\documentclass[preprint,aps,12pt,showpacs,nofootinbib,tightenlines]{revtex4}
\usepackage{mathrsfs}
\usepackage{amsmath}
\usepackage{amssymb}
\usepackage{epsfig}
\usepackage{epstopdf}
\usepackage{graphicx}
\usepackage{subfigure}
\usepackage{booktabs}
\usepackage{float}
\usepackage{amsmath}
\usepackage{color}
\usepackage{multirow}

\def\be{\begin{eqnarray}}
\def\en{\end{eqnarray}}
\def\non{\nonumber\\}

\def\prd{{Phys. Rev. D}~}

\begin{document}
%%--------------------------------------------
\title{
 Semileptonic $B_{c}$ meson decays to S-wave charmonia and $X(3872)$ within the covariant light-front approach  }
\author{Zhi-Jie Sun$^1$, Si-Yang Wang$^2$, Zhi-Qing Zhang$^1$
\footnote{ zhangzhiqing@haut.edu.cn (corresponding author)}, You-Ya Yang$^1$ and Zi-Yu Zhang$^1$ } %%
\affiliation{ \it \small $^1$  Institute of Theoretical Physics, School of Sciences, Henan University of Technology,
 Zhengzhou, Henan 450052, China; \\
\small $^2$ \it
Institute of Particle Physics and Key Laboratory of Quark and Lepton Physics (MOE),\\
\small \it Central China Normal University, Wuhan, Hubei 430079, China }
\date{\today}
\begin{abstract}
In this work, we investigate the semileptonic decays of $B_{c}$ meson to $\eta_{c}(1S,2S,3S)$, $\psi(1S,2S,3S)$ and $X(3872)$ within the framework of covariant light-front quark model (CLFQM). We combine the helicity amplitudes via the corresponding form factors to obtain the branching ratios of the semileptonic decays $B_{c}\to \eta_{c}(1S, 2S, 3S)\ell\nu_{\ell}$, $B_{c}\to \psi(1S, 2S, 3S))\ell\nu_{\ell}$ and $B_{c}\to X(3872)\ell\nu_{\ell}$ with $\ell=e,\mu,\tau$. In view of the $R_{J/\Psi}$ anomaly released by the LHCb collaboration, it is necessary to calculate the ratios $R_X$ with $X=\psi(1S,2S,3S),\eta_c(1S,2S,3S),X(3872)$ systematically, which are helpful to check the lepton flavor universality (LFU).
% These results
%In 2017, the LHCb collaboration  released an experimental observation value $R_{J / \psi}|_{\exp }$, which was larger than the theoretical %prediction value 2$\sigma$ of the Standard Model(SM).
Furthermore, we also take into account another two physical observables, one is the longitudinal polarization fraction $f_{L}$ and the other is the forward-backward asymmetry $A_{FB}$, which can provide new clues to understand the $R_{J /\Psi}$ anomaly. Such theoretical predictions are necessary and interesting, which can be tested in the future LHCb experiments.
\end{abstract}

 \vspace{1cm}

\maketitle

%=======================================================================
%                     Introduction
%=======================================================================
\section{Introduction}\label{intro}
In this paper we research the exclusive semileptonic $B_c$ decays to $\eta_{c}(1S,2S,3S)$, $\psi(1S,2S,3S)$ and $X(3872)$ by using the
the covariant Light-Front quark model (CLFQM).
%extend our previous study in the two-body  nonleptonic of $B_{c}$ meson to charmonia and charmed mesons\cite{Zhang:2023ypl}
The traditional light-front quark model (LFQM) was initially proposed by Terentev and Berestesky \cite{tere,bere}, which is based on the
light-front formalism of Hamiltonian dynamics \cite{DiracO}, and later developed and applied to determinate the transition form factor, decay constant and distribution amplitude \cite{jaus0,jaus1,Choi}.  Unfortunately, the Lorentz covariance of the matrix
element is lost and the zero-mode contributions can not be handled in the traditional LFQM. To compensate for these deficiencies, Jaus put forward the CLFQM \cite{{jaus}}, where the spurious contribution being
dependent on the orientation of the light-front can be eliminated by the inclusion of the zero-mode contributions.  This model
has been used successfully to investigate the nonleptonic and simileptonic $B_{(c)}$ meson decays \cite{Zhang:2023ypl,Wang09,Ke14,hycheng,Wang:2009mi}.

The semileptonic $B_{c}$ meson decays to charmonium states play an important role in testing the Standard Model (SM) and searching for new physics (NP), for example,  the issues about test of the lepton flavor universality (LFU), determination of the CKM matrix element $V_{cb}$ are often searched by experiments and theories. In 2017, the ratio of the semileptonic branching fractions $R_{J / \psi}|_{\exp }$ was measured by the LHCb Collaboration \cite{LHCb:2017vlu} ,
\be
\label{rjpsi}
\left.R_{J / \psi}\right|_{\exp }=\frac{\mathcal{B} r\left(B_{c}^{+} \rightarrow J / \psi \tau^{+} \nu_{\tau}\right)}{\mathcal{B}r\left(B_{c}^{+} \rightarrow J / \psi \mu^{+} \nu_{\mu}\right)}=0.71 \pm 0.17 \pm 0.18,
\en
which lies within 2$\sigma$ above the range of existing SM predictions  \cite{Cohen:2018dgz,AA}. It was considered as one of the most fascinating puzzles in flavor physics in recent years. In order to provide a generalized and complementary check, it is useful to measure the values of
$R_{\psi(2S,3S)}$ and $R_{\eta_{c}(1S,2S,3S)}$. At present, there also exist some results about these ratios predicted by other approaches. Furthermore, another two physical measurements are also sensitive to NP, one is the  longitudinal polarization fraction $f_{L}$ and the other is the forward-backward asymmetry $A_{FB}$. These observations can be represented by the helicity amplitudes, which are combined via the corresponding form factors.

Many different theoretical methods have been devoted to studying the semileptonic $B_{c}$ meson decays to charmonium states, such as the nonrelativistic QCD (NRQCD) \cite{C.F131,Qiao:132}, the Bethe-Salpeter (BS) method \cite{R.R,ZZT}, the relativistic quark model (RQM)\cite{Ebert10,Ebert101,Ivanov06}, the light-cone QCD sum rules\cite{Huang07,M08},  the relativistic constituent quark model (RCQM) \cite{Anisimov:1998uk,Anisimov:1998uk1}, the non-relativistic quark model (NRQM)\cite{Hernndez06}, the QCD potential model(QCDPM)\cite{P00}, the Isgur-Scora-Grinstein-Wise (ISGW2) quark model\cite{I}, the perturbative QCD (PQCD) approach \cite{Rui:2016opu,Wang:2012lrc,JFSD}, the covariant confined quark model (CCQM) \cite{Ivanov:2000aj}, the QCD sum rules (QCDSR) \cite{Kiselev:2000pp}, the covariant quark model (CQM)\cite{Issadykov:2017wlb}, the relativistic independent quark (RIQ) model\cite{Nayak:2022gdo}, and so on.

This paper is organized as follows. In section \ref{form}, the formalism of the CLFQM and the helicity amplitudes combined via form factors are presented. Numerical results for the branching ratios, the longitudinal polarization fractions $f_{L}$ and the forward-backward asymmetries $A_{FB}$ for these semileptonic $B_c$ decays are listed in section \ref{numer}. Detailed comparisons with other theoretical values and relevant discussions are also included. The summary is presented in section \ref{sum}. Some specific rules when performing the $p^-$ integration and analytical expressions of the $B_c\to \eta_c(1S,2S,3S), \psi(1S,2S,3S), X(3872)$ transition form factors are collected in Appendix A and B, respectively.

\section{Formalism}\label{form}
\subsection{Covariant Light-Front Quark Model}
\begin{figure}[H]
\centering \subfigure{
\begin{minipage}{5cm}
\centering
\includegraphics[width=5cm]{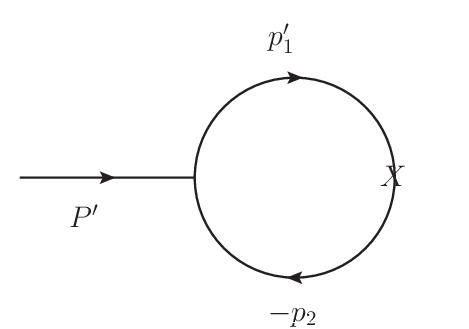}
\end{minipage}}
\subfigure{
\begin{minipage}{6cm}
\centering
\includegraphics[width=6cm]{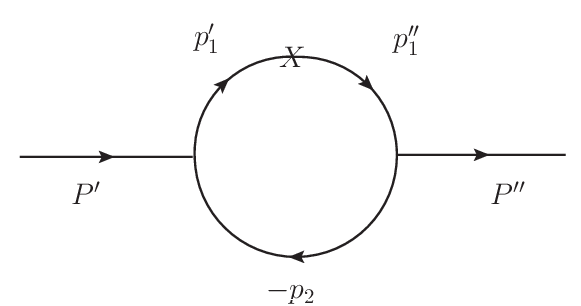}
\end{minipage}}
\caption{Feynman diagrams for $B_c$ decay (left) and transition
(right) amplitudes, where $P^{\prime(\prime\prime)}$ is the
incoming (outgoing) meson momentum, $p^{\prime(\prime\prime)}_1$
is the quark momentum, $p_2$ is the anti-quark momentum. The X in the diagrams denotes the vector or axial-vector transition vertex.}
\label{feyn}
\end{figure}
In the covariant light-front quark model, we will employ the light-front decomposition of the momentum  $p=(p^-,p^+,p_\perp)$ with
$p^\pm=p^0\pm p_z, p^2=p^+p^--p^2_\perp$.
The Feynman diagrams for $B_c$ meson decay and transition amplitudes are shown in Fig. \ref{feyn}.
The incoming (outgoing) meson has the mass $M^\prime(M^{\prime\prime})$
with the momentum $P^\prime=p_1^\prime+p_2 (P^{\prime\prime}=p_1^{\prime\prime}+p_2)$, where $p_{1}^{\prime(\prime\prime)} $
and $p_{2}$ are the momenta of the quark and anti-quark
inside the incoming (outgoing) meson with the mass $m_{1}^{\prime(\prime\prime)}$and $m_{2}$, respectively. Here we use the same notations
as those in Refs. \cite{jaus,hycheng} and $M^\prime$ refers to $m_{B_c}$ for $B_c$ meson decays.
These momenta can be expressed in terms of the internal variables $(x_{i},p{'}_{\perp})$ as
\be
p_{1,2}^{\prime+}=x_{1,2} P^{\prime+}, \quad p_{1,2 \perp}^{\prime}=x_{1,2} P_{\perp}^{\prime} \pm p_{\perp}^{\prime},
\en
where $x_{1}+x_{2}=1$. Using these internal variables,
we can define some quantities for the incoming meson which will be used in the following calculations
\be
M_{0}^{\prime 2} &=&\left(e_{1}^{\prime}+e_{2}\right)^{2}=\frac{p_{\perp}^{\prime 2}+m_{1}^{\prime 2}}{x_{1}}
+\frac{p_{\perp}^{2}+m_{2}^{2}}{x_{2}}, \quad \widetilde{M}_{0}^{\prime}=\sqrt{M_{0}^{\prime 2}-\left(m_{1}^{\prime}-m_{2}\right)^{2}},\non
e_{i}^{(\prime)} &=&\sqrt{m_{i}^{(\prime) 2}+p_{\perp}^{\prime 2}+p_{z}^{\prime 2}}, \quad \quad p_{z}^{\prime}
=\frac{x_{2} M_{0}^{\prime}}{2}-\frac{m_{2}^{2}+p_{\perp}^{\prime 2}}{2 x_{2} M_{0}^{\prime}}, \label{m0p}
\en
where $M'_0$ is the kinetic invariant mass of the incoming meson and can be expressed as the energies of the quark and the anti-quark
$e^{(\prime)}_i$. It is similar to the case of the outgoing meson.
To calculate the amplitudes for the transition form factors,
we need the Feynman rules for the meson-quark-antiquark vertices ($i\Gamma'_M(M=\eta_c,\psi, X)$), which are listed in Tab. \ref{tab0}.
Here $X(3872)$ is considered as a $1^{++}$ charmonium sate in our calculations \cite{zhang} and replaced with $X$ in some places for simply.
\begin{table}
\caption{Feynman rules for the vertices $i\Gamma^\prime_M$ of the incoming meson-quark-antiquark, where $p^\prime_1$ and $p_2$ are the
quark and antiquark momenta, respectively. It is noticed that for the outgoing meson, we should use $i(\gamma_0\Gamma^{\prime\dag}_M\gamma_0)$ for the relevant vertices.}
\begin{center}
\begin{tabular}{c|c}
\hline\hline $M\left({ }^{2 S+1} L_{J}\right)$ & $i \Gamma_{M}^{\prime}$ \\
\hline
$\eta_c\left({ }^{1} S_{0}\right) $&$H_{\eta_c}^{\prime} \gamma_{5}$\\
$\psi\left({ }^{3} S_{1}\right) $&$ i H_{\psi}^{\prime}\left[\gamma_{\mu}-\frac{1}{W_{\psi}^{\prime}}\left(p_{1}^{\prime}-p_{2}\right)_{\mu}\right]$\\
$X\left({ }^{3} P_{1}\right) $&$-i H_{X}^{\prime}\left[\gamma_{\mu}+\frac{1}{W_{X}^{\prime}}\left(p_{1}^{\prime}-p_{2}\right)_{\mu}\right] \gamma_{5}$\\
 \hline\hline
\end{tabular}\label{tab0}
\end{center}
\end{table}
\subsection{Wave Functions and Decay Constants}\label{decaycons}
In order to calculate the form factors, we need to specify the light-front wave functions. In principle, one can obtain them by solving the
relativistic Schr$\ddot{o}$dinger equation. But it is difficult to obtain the exact solution in many cases. Usually, the Gaussian-type and
harmonic-oscillator-type
wave functions are used to describe various hadronic structures in many works.
In the present work, we shall use the phenomenological Gaussian-type wave functions
\be
\varphi^{\prime} &=&\varphi^{\prime}\left(x_{2}, p_{\perp}^{\prime}\right)=4\left(\frac{\pi}{\beta^{\prime 2}}\right)^{\frac{3}{4}}
\sqrt{\frac{d p_{z}^{\prime}}{d x_{2}}} \exp \left(-\frac{p_{z}^{\prime 2}+p_{\perp}^{\prime 2}}{2 \beta^{\prime 2}}\right),\non
\varphi_{p}^{\prime} &=&\varphi_{p}^{\prime}\left(x_{2}, p_{\perp}^{\prime}\right)=\sqrt{\frac{2}{\beta^{\prime 2}}} \varphi^{\prime},
\quad \frac{d p_{z}^{\prime}}{d x_{2}}=\frac{e_{1}^{\prime} e_{2}}{x_{1} x_{2} M_{0}^{\prime}},\label{betap}
\en
where the parameter $\beta'$ describes the momentum distribution and is about of order $\Lambda_{QCD}$.
It is usually determined by the decay constants through the analytic expressions in the conventional light-front approach, which
are given as follows \cite{jaus,hycheng}
\be
f_{\eta_c}&=&\frac{N_{c}}{16 \pi^{3}} \int d x_{2} d^{2} p_{\perp}^{\prime} \frac{h^\prime_{\eta_c}}{x_{1} x_{2} (M^{\prime2}-M^{\prime2}_0)}
4\left(m_{1}^{\prime} x_{2}+m_{2} x_{1}\right),\\
f_{\psi}&=&\frac{N_{c}}{4 \pi^{3} M^{\prime}} \int d x_{2} d^{2} p_{\perp}^{\prime} \frac{h_{\psi}^{\prime}}{x_{1} x_{2}\left(M^{\prime 2}-M_{0}^{\prime 2}\right)}\non &&
\times\left[x_{1} M_{0}^{\prime 2}-m_{1}^{\prime}\left(m_{1}^{\prime}-m_{2}\right)-p_{\perp}^{\prime 2}+\frac{m_{1}^{\prime}+m_{2}}{w_{\psi}^{\prime}} p_{\perp}^{\prime 2}\right],\\
f_{X}&=&-\frac{N_{c}}{4 \pi^{3} M^{\prime}}  \int d x_{2} d^{2} p_{\perp}^{\prime} \frac{h_{X}^{\prime}}{x_{1} x_{2}\left(M^{\prime 2}-M_{0}^{\prime 2}\right)}\non &&
\times\left[x_{1} M_{0}^{\prime 2}-m_{1}^{\prime}\left(m_{1}^{\prime}+m_{2}\right)-p_{\perp}^{\prime 2}-\frac{m_{1}^{\prime}-m_{2}}{w_{X}^{\prime}} p_{\perp}^{\prime 2}\right],
\en
where $m_{1}^{\prime}$ and $m_{2}$ are the constituent quarks of meson $M (M=\eta_c,\psi,X)$, and $M_{0}^{\prime}$ is defined in Eq. (\ref{m0p}).
The explicit forms of $h'_{M}$  are given by \cite{hycheng}
\be
h_{\eta_c}^{\prime} &=&h_{\psi}^{\prime}=\left(M^{\prime 2}-M_{0}^{\prime 2}\right) \sqrt{\frac{x_{1} x_{2}}{N_{c}}} \frac{1}{\sqrt{2} \widetilde{M}_{0}^{\prime}} \varphi^{\prime},\label{hp}\\
\sqrt{\frac{2}{3}} h_{X}^{\prime}&=&\left(M^{\prime 2}-M_{0}^{\prime 2}\right) \sqrt{\frac{x_{1} x_{2}}{N_{c}}} \frac{1}{\sqrt{2} \widetilde{M}_{0}^{\prime}}
\frac{\widetilde{M}_{0}^{\prime 2}}{2 \sqrt{3} M_{0}^{\prime}} \varphi_{p}^{\prime},\label{hs3a}
\en
where $\varphi^{\prime}$ and $\varphi_{p}^{\prime}$ are the light-front momentum distribution amplitudes for the S-wave and P-wave mesons, respectively, defined in Eq. (\ref{betap}).
\subsection{Helicity amplitudes and Observables  }

 Since the form factors involving the fitted parameters for the $B_{c}\to \eta_{c}(1S,2S,3S)$, $B_{c}\to \psi(1S,2S,3S)$ and $B_{c}\to X(3872)$ transitions have been investigated  in our recent work \cite{Zhang:2023ypl}, so it is convenient to obtain the differential decay widths of these semileptontic $B_c$ decays by the combination of the helicity amplitudes via form factors, which are listed as following
 \begin{footnotesize}
\begin{eqnarray}
 \frac{d\Gamma(B_c\to \eta_c\ell\nu_\ell)}{dq^2} &=&(\frac{q^2-m_\ell^2}{q^2})^2\frac{ {\sqrt{\lambda(m_{B_c}^2,m_{\eta_c}^2,q^2)}} G_F^2 |V_{cb}|^2} {384m_{B_c}^3\pi^3}
 \times \frac{1}{q^2} \nonumber\\
 &&\;\;\;\times \left\{ (m_\ell^2+2q^2) \lambda(m_{B_c}^2,m_{\eta_c}^2,q^2) F_1^2(q^2)  +3 m_\ell^2(m_{B_c}^2-m_{\eta_c}^2)^2F_0^2(q^2)
 \right\},\label{eq:pp}\\
 \frac{d\Gamma_L(B_c\to \psi\ell\nu_\ell)}{dq^2}&=&(\frac{q^2-m_\ell^2}{q^2})^2\frac{ {\sqrt{\lambda(m_{B_c}^2,m_\psi^2,q^2)}} G_F^2 |V_{cb}|^2} {384m_{B_c}^3\pi^3}
 \times \frac{1}{q^2} \left\{ 3 m_\ell^2 \lambda(m_{B_c}^2,m_\psi^2,q^2) A_0^2(q^2)\right.+(m_\ell^2+2q^2)\nonumber\\
 && \times\left.\left|\frac{1}{2m_\psi}  \left[
 (m_{B_c}^2-m_\psi^2-q^2)(m_{B_c}+m_\psi)A_1(q^2)-\frac{\lambda(m_{B_c}^2,m_\psi^2,q^2)}{m_{B_c}+m_\psi}A_2(q^2)\right]\right|^2
 \right\},\label{eq:decaywidthlon}\;\;\;\;\;\;
 \end{eqnarray}
\end{footnotesize}
\begin{footnotesize}
\begin{eqnarray}
\frac{d\Gamma_\pm(B_c\to
 \psi\ell\nu_\ell)}{dq^2}&=&(\frac{q^2-m_\ell^2}{q^2})^2\frac{ {\sqrt{\lambda(m_{B_c}^2,m_\psi^2,q^2)}} G_F^2 |V_{cb}|^2} {384m_{B_c}^3\pi^3}
  \nonumber\\
 &&\;\;\times \left\{ (m_\ell^2+2q^2) \lambda(m_{B_c}^2,m_\psi^2,q^2)\left|\frac{V(q^2)}{m_{B_c}+m_\psi}\mp
 \frac{(m_{B_c}+m_\psi)A_1(q^2)}{\sqrt{\lambda(m_{B_c}^2,m_\psi^2,q^2)}}\right|^2
 \right\},\label{eq:widthlon2}\\
 \frac{d\Gamma_L(B_c\to X\ell\nu_\ell)}{dq^2}&=&(\frac{q^2-m_\ell^2}{q^2})^2\frac{ {\sqrt{\lambda(m_{B_c}^2,m_X^2,q^2)}} G_F^2 |V_{cb}|^2} {384m_{B_c}^3\pi^3}
 \times \frac{1}{q^2} \left\{ 3 m_\ell^2 \lambda(m_{B_c}^2,m_X^2,q^2) V_0^2(q^2)+(m_\ell^2+2q^2)\right.\nonumber\\
 &&\times  \left. \left|\frac{1}{2m_X}  \left[
 (m_{B_c}^2-m_X^2-q^2)(m_{B_c}-m_X)V_1(q^2)-\frac{\lambda(m_{B_c}^2,m_X^2,q^2)}{m_{B_c}-m_X}V_2(q^2)\right]\right|^2
 \right\},\label{eq:decaywidthlon}\\
 \frac{d\Gamma_\pm(B_c\to
 X\ell\nu_\ell)}{dq^2}&=&(\frac{q^2-m_\ell^2}{q^2})^2\frac{ {\sqrt{\lambda(m_{B_c}^2,m_X^2,q^2)}} G_F^2 |V_{cb}|^2} {384m_{B_c}^3\pi^3}
  \nonumber\\
 &&\;\;\times \left\{ (m_\ell^2+2q^2) \lambda(m_{B_c}^2,m_X^2,q^2)\left|\frac{A(q^2)}{m_{B_c}-m_X}\mp
 \frac{(m_{B_c}-m_X)V_1(q^2)}{\sqrt{\lambda(m_{B_c}^2,m_X^2,q^2)}}\right|^2
 \right\},
\end{eqnarray}
\end{footnotesize}
where $\lambda(q^2)=\lambda(m^{2}_{B_{c}},m^{2}_{\eta_c(\psi,X)},q^{2})=(m^{2}_{B_{c}}+m^{2}_{\eta_c(\psi,X)}-q^{2})^{2}-4m^{2}_{B_{c}}m^{2}_{\eta_c(\psi,X)}$ and $m_{\ell}$ is the mass of the lepton $\ell$ with $\ell=e,\mu,\tau$\footnote{For now on, we use $\ell$ to represent $e,\mu,\tau$ and use $\ell^\prime$ to represent $e,\mu$ for simplicity. }. It is noticed that although the electron and nuon are very light compared with the charm quark, we do not ignore their masses in our calculations.
The combined transverse and total differential decay widths are defined as
\be
\frac{d \Gamma_{T}}{d q^{2}}=\frac{d \Gamma_{+}}{d q^{2}}+\frac{d \Gamma_{-}}{d q^{2}}, \quad \frac{d \Gamma}{d q^{2}}=\frac{d \Gamma_{L}}{d q^{2}}+\frac{d \Gamma_{T}}{d q^{2}}.
\en

For $\psi(1S,2S,3S)$ and $X(3872)$, it is meaningful to define the polarization fraction due to the existence of different polarizations
\be
f_{L}=\frac{\Gamma_{L}}{\Gamma_{L}+\Gamma_{+}+\Gamma_{-}}. \label{eq:fl}
\en
As to the forward-backward asymmetry, the analytical expression is defined as \cite{ptau3},
\be
A_{FB} = \frac{\int^1_0 {d\Gamma \over dcos\theta} dcos\theta - \int^0_{-1} {d\Gamma \over dcos\theta} dcos\theta}
{\int^1_{-1} {d\Gamma \over dcos\theta} dcos\theta} = \frac{\int b_\theta(q^2) dq^2}{\Gamma_{B_{c}}},\label{eq:AFB}
\en
where $\theta$ is the angle between the 3-momenta of the lepton $\ell$ and the initial $B_c$ meson in the $\ell\nu$ rest frame. The function $b_{\theta}(q^2)$ represents the angular coefficient, which can be written as \cite{ptau3}
\be
b_\theta^{(\eta_c)}(q^2) &=& {G_F^2 |V_{cb}|^2 \over 128\pi^3 m_{B_{c}}^3} q^2 \sqrt{\lambda(q^2)}
\left( 1 - {m_\ell^2 \over q^2} \right)^2 {m_\ell^2 \over q^2} ( H^s_{V,0}H^s_{V,t} ) , \label{eq:btheta1}\\
b_\theta^{(\psi,X)}(q^2) &=& {G_F^2 |V_{cb}|^2 \over 128\pi^3 m_{B_{c}}^3} q^2 \sqrt{\lambda(q^2)}
\left( 1 - {m_\ell^2 \over q^2} \right)^2 \left[ {1 \over 2}(H_{V,+}^2-H_{V,-}^2)+ {m_\ell^2 \over q^2} ( H_{V,0}H_{V,t} ) \right],
\label{eq:btheta2}
\en
where the helicity amplitudes
\be
H^s_{V,0}\left(q^{2}\right)  =\sqrt{\frac{\lambda\left(q^{2}\right)}{q^{2}}} F_{1}\left(q^{2}\right),
H^s_{V,t}\left(q^{2}\right)  =\frac{m_{B_{c}}^{2}-m_{{\eta_c}}^{2}}{\sqrt{q^{2}}} F_{0}\left(q^{2}\right),
\en
for the $B_c$ to $\eta_c(1S,2S,3S)$ meson transitions, and
the helicity amplitudes
\be
H_{V,\pm}\left(q^{2}\right)&=&\left(m_{B_{c}}+{m_\psi}\right) A_{1}\left(q^{2}\right) \mp \frac{\sqrt{\lambda\left(q^{2}\right)}}{m_{B_{c}}+m_{\psi}} V\left(q^{2}\right), \non
H_{V,0}\left(q^{2}\right)&=&\frac{m_{B_{c}}+m_{\psi}}{2 m_{\psi} \sqrt{q^{2}}}\left[-\left(m_{B_{c}}^{2}-m_{\psi}^{2}-q^{2}\right) A_{1}\left(q^{2}\right)+\frac{\lambda\left(q^{2}\right) A_{2}\left(q^{2}\right)}{\left(m_{B_{c}}+m_{\psi}\right)^{2}}\right],\non
H_{V,t}\left(q^{2}\right)&=&-\sqrt{\frac{\lambda\left(q^{2}\right)}{q^{2}}} A_{0}\left(q^{2}\right),
\en
for the $B_{c}$ to $\psi(1S,2S,3S)$ transitions, and
the helicity amplitudes
\be
H_{V,\pm}\left(q^{2}\right)&=&\left(m_{B_{c}}-{m_X}\right) V_{1}\left(q^{2}\right) \mp \frac{\sqrt{\lambda\left(q^{2}\right)}}{m_{B_{c}}-m_{X}} A\left(q^{2}\right), \non
H_{V,0}\left(q^{2}\right)&=&\frac{m_{B_{c}}-m_X}{2 m_{X} \sqrt{q^{2}}}\left[-\left(m_{B_{c}}^{2}-m_{X}^{2}-q^{2}\right) V_{1}\left(q^{2}\right)+\frac{\lambda\left(q^{2}\right) V_{2}\left(q^{2}\right)}{\left(m_{B_{c}}-m_{X}\right)^{2}}\right],\non
H_{V,t}\left(q^{2}\right)&=&-\sqrt{\frac{\lambda\left(q^{2}\right)}{q^{2}}} V_{0}\left(q^{2}\right),
\en
for the $B_{c}$ to $X(3872)$ transition. It is noticed that the subscript $V$ in each helicity amplitude refers to the $\gamma_\mu(1-\gamma_5)$ current.
The transition form factors are collected in Appendix B.

\section{Numerical results and discussions} \label{numer}
The adopted input parameters \cite{pdg}, such as the constituent quark masses, the hadron and lepton masses, the $B_c$ meson life and the Cabibbo-Kobayashi-Maskawa (CKM) matrix element $V_{cb}$, in our numerical calculations are listed in Table \ref{tab:constant}.
In the calculations of the helicity amplitudes, the transition form factors are the most important inputs, which have been calculated in our previous work \cite{Zhang:2023ypl}.  Those
parameterized form factors are extrapolated from the space-like region to the  time-like region  by using following expression,
 \begin{table}[H]
\caption{The values of the input parameters.}
\label{tab:constant}
\begin{tabular*}{16.5cm}{@{\extracolsep{\fill}}l|cccccc}
  \hline\hline
\textbf{Mass(\text{GeV})} &$m_{b}=4.8$
&$m_{c}=1.4$&$m_{e}=0.000511$&$m_{\mu}=0.106$&$m_{\tau}=1.78$   \\[1ex]
& $ m_{J/\psi}=3.0969$  & $m_{\psi(2S)}=3.6861$& $m_{\psi(3S)}=4.039 $& $ M_{B_c}=6.27447$ \\[1ex]
& $m_{\eta_c}=2.9839$ &$ m_{\eta_c(2S)}=3.6375$ & $m_{\eta_c(3S)}=3.940$&$ m_{X(3872)}=3.87165$ \\[1ex]
\hline
\end{tabular*}
\begin{tabular*}{16.5cm}{@{\extracolsep{\fill}}l|c|c|cc}
  \hline
{{\textbf{CKM}}} &$V_{cb}=(40.8\pm1.4) \times 10^{-3}$& \textbf{Lifetime}&$\tau_{B_c}=(0.510\pm0.009)\times 10^{-12}\text{s}$ \\[1ex]
\hline\hline
\end{tabular*}
\end{table}

\begin{eqnarray}
 F(q^2)=F(0){\rm exp}(a \frac{q^2}{m_{B_c}^2}+b(\frac{q^2}{m_{B_c}^2})^2)
\end{eqnarray}
where $F(q^{2})$ denotes different form factors. The corresponding results are listed in Tab. \ref{btopv}.
\begin{table}[H]
\caption{$B_c\to \eta_c{(1S,2S,3S)}, \psi(1S,2S,3S), X(3872)$ form factors in the CLFQM.
The uncertainties are from the decay constants of $B_c$ and the final state mesons.}
\begin{center}
\scalebox{0.8}{
\begin{tabular}{ccc||cccc}
\hline\hline
$$&$B_{c}\rightarrow \eta_{c}$&$$& $$&$B_{c}\rightarrow J/\psi$&$$\\
\hline
$$&$F_{1}$& $F_{0}$&$V$&$A_{0}$&$A_{1}$&$A_{2}$\\
\hline
$F(0)$&$0.60^{+0.00+0.01}_{-0.00-0.01}$&$0.60^{+0.00+0.01}_{-0.01-0.00}$&$0.76^{+0.00+0.04}_{-0.00-0.04}$&$0.55^{+0.00+0.03}_{-0.00-0.04}$&$0.53^{+0.00+0.02}_{-0.03-0.00}$&$0.49^{+0.00+0.00}_{-0.00-0.01}$\\
$F(q^{2}_{max})$&$1.06^{+0.00+0.03}_{-0.00-0.03}$&$0.85^{+0.00+0.02}_{-0.01-0.02}$&$1.37^{+0.00+0.11}_{-0.00-0.10}$&$0.76^{+0.00+0.06}_{-0.00-0.07}$&$0.78^{+0.01+0.02}_{-0.01-0.05}$&$0.84^{+0.00+0.03}_{-0.00-0.00}$\\
$a$& $1.95^{+0.01+0.03}_{-0.01-0.03}$&$1.44^{+0.00+0.03}_{-0.00-0.03}$&$2.16^{+0.01+0.09}_{-0.01-0.08}$&$1.22^{+0.02+0.07}_{-0.02-0.07}$&$1.45^{+0.03+0.09}_{-0.01-0.09}$&$1.97^{+0.01+0.11}_{-0.01-0.11}$\\
$b$& $0.48^{+0.00+0.01}_{-0.00-0.01}$&$-0.62^{+0.02+0.02}_{-0.02-0.03}$&$0.53^{+0.00+0.01}_{-0.00-0.01}$&$0.16^{+0.00+0.00}_{-0.00-0.00}$&$0.29^{+0.00+0.02}_{-0.00-0.00}$&$0.43^{+0.00+0.03}_{-0.00-0.03}$\\
\hline
$$&$B_{c}\rightarrow \eta_{c}(2S)$&$$& $$&$B_{c}\rightarrow \psi(2S)$&$$\\
\hline
$$&$F_{1}$& $F_{0}$&$V$&$A_{0}$&$A_{1}$&$A_{2}$\\
\hline
$F(0)$&$0.37^{+0.00+0.12}_{-0.00-0.18}$&$0.37^{+0.00+0.12}_{-0.00-0.18}$&$0.57^{+0.00+0.01}_{-0.00-0.00}$&$0.41^{+0.00+0.00}_{-0.00-0.00}$&$0.35^{+0.00+0.00}_{-0.00-0.00}$&$0.17^{+0.00+0.00}_{-0.00-0.00}$\\
$F(q^{2}_{max})$&$0.48^{+0.00+0.28}_{-0.00-0.31}$&$0.41^{+0.00+0.28}_{-0.01-0.28}$&$0.67^{+0.00+0.02}_{-0.00-0.00}$&$0.44^{+0.00+0.00}_{-0.00-0.00}$&$0.35^{+0.00+0.00}_{-0.00-0.00}$&$0.12^{+0.00+0.00}_{-0.00-0.00}$\\
$a$& $1.44^{+0.00+0.92}_{-0.00-0.66}$&$0.73^{+0.01+0.99}_{-0.01-0.95}$&$1.01^{+0.01+0.01}_{-0.01-0.02}$&$0.39^{+0.01+0.01}_{-0.01-0.01}$&$0.08^{+0.01+0.02}_{-0.02-0.03}$&$-1.53^{+0.07+0.09}_{-0.09-0.13}$\\
$b$&$0.15^{+0.02+0.50}_{-0.02-0.34}$&$-0.81^{+0.02+0.34}_{-0.02-0.28}$&$-0.16^{+0.03+0.01}_{-0.03-0.02}$&$-0.15^{+0.02+0.01}_{-0.02-0.01}$&$-0.69^{+0.03+0.01}_{-0.04-0.02}$&$-3.67^{+0.14+0.13}_{-0.19-0.21}$\\
\hline
$$&$B_{c}\rightarrow \eta_{c}(3S)$&$$& $$&$B_{c}\rightarrow \psi(3S)$&$$\\
\hline
$$&$F_{1}$& $F_{0}$&$V$&$A_{0}$&$A_{1}$&$A_{2}$\\
\hline
$F(0)$&$0.29^{+0.00+0.04}_{-0.00-0.05}$&$0.29^{+0.00+0.04}_{-0.00-0.05}$&$0.46^{+0.00+0.02}_{-0.00-0.02}$&$0.31^{+0.00+0.01}_{-0.00-0.01}$&$0.27^{+0.00+0.01}_{-0.00-0.01}$&$0.14^{+0.00+0.01}_{-0.00-0.01}$\\
$F(q^{2}_{max})$&$0.36^{+0.00+0.07}_{-0.00-0.08}$&$0.32^{+0.00+0.07}_{-0.00-0.08}$&$0.53^{+0.00+0.03}_{-0.00-0.03}$&$0.33^{+0.00+0.01}_{-0.00-0.01}$&$0.28^{+0.00+0.02}_{-0.00-0.01}$&$0.12^{+0.00+0.02}_{-0.00-0.02}$\\
$a$&$1.53^{+0.00+0.29}_{-0.00-0.23}$&$0.85^{+0.01+0.44}_{-0.01-0.31}$&$1.14^{+0.03+0.04}_{-0.03-0.03}$&$0.49^{+0.03+0.03}_{-0.02-0.03}$&$0.25^{+0.02+0.04}_{-0.02-0.03}$&$-1.01^{+0.02+0.04}_{-0.02-0.04}$\\
$b$&$0.23^{+0.01+0.13}_{-0.01-0.13}$&$-0.74^{+0.01+0.05}_{-0.00-0.24}$&$-0.01^{+0.01+0.01}_{-0.01-0.01}$&$-0.04^{+0.01+0.01}_{-0.01-0.01}$&$-0.45^{+0.01+0.00}_{-0.01-0.00}$&$-2.73^{+0.01+0.00}_{-0.01-0.00}$\\
\hline
$$&$B_{c}\rightarrow X(3872)$&$$&$A$& $V_{0}$&$V_{1}$&$V_{2}$\\
\hline
$$&$F(0)$&$$&$0.28^{+0.00+0.02}_{-0.00-0.03}$&$0.21^{+0.00+0.01}_{-0.00-0.01}$&$1.13^{+0.00+0.01}_{-0.00-0.03}$&$0.11^{+0.00+0.01}_{-0.01-0.01}$\\
$$&$F(q^{2}_{max})$&$$&$0.37^{+0.00+0.03}_{-0.00-0.04}$&$0.19^{+0.00+0.02}_{-0.00-0.02}$&$1.10^{+0.00+0.05}_{-0.00-0.06}$&$0.12^{+0.00+0.01}_{-0.01-0.01}$\\
$$&$a$&$$&$1.85^{+0.02+0.09}_{-0.02-0.08}$&$-0.52^{+0.01+0.38}_{-0.01-0.32}$&$-0.05^{+0.01+0.24}_{-0.01-0.20}$&$0.77^{+0.03+0.04}_{-0.03-0.04}$\\
$$&$b$&$$&$0.38^{+0.01+0.01}_{-0.01-0.03}$&$-1.45^{+0.02+0.36}_{-0.03-0.32}$&$-1.03^{+0.00+0.15}_{-0.00-0.12}$&$-0.61^{+0.02+0.08}_{-0.02-0.12}$\\
 \hline\hline
\end{tabular}\label{btopv}
}
\end{center}
\end{table}

The branching ratios for these semileptonic $B_c$ decays to the S-wave ground charmonium states are collected in Tab. \ref{tab1q}. The uncertainties arise from the $B_{c}$ meson lifetime, the decay constants of $B_{c}$ and final state mesons, respectively. Obviously, because the mass of $\tau$ lepton is much larger than those of $e, \mu$ leptons, $\mathcal{B} r(B^+_c\to \eta_c(J/\Psi)\ell^{\prime+}\nu_{\ell^\prime})$ are about $3\sim 4$ times larger than $\mathcal{B} r(B^+_c\to \eta_c(J/\Psi)\tau^+\nu_\tau)$. For comparison, we also list the results calculated by other approaches.  The predictions given by the NRQCD \cite{C.F131,Qiao:132} and the PQCD approach \cite{Rui:2016opu} are much larger than others. It is because that the former involves the large QCD correction K factor and the NLO charmonium wave functions, and the latter includes the large weak transition form factors. These differences can be clarified by the future LHCb experiments. Certainly, our predictions are consistent with most of other theoretical results, such as the BS equation \cite{R.R}, the RQM \cite{Ebert10,Ebert101,Ivanov06}, the RCQM \cite{Anisimov:1998uk,Anisimov:1998uk1}, the NRQM \cite{Hernndez06}, the QCDPM \cite{P00},
the CQM \cite{Issadykov:2017wlb} and so on.

\begin{table}[H]
\caption{Branching ratios (in\%) of the semileptonic $B_{c}$ decays to $\eta_c$ and $J/\Psi$.}
\begin{center}
\scalebox{0.7}{
\begin{tabular}{c|c|c|c|c|c|c}
\hline\hline
  $$&$ \mathcal{B} r(B_{c}^{+}\to \eta_{c} e^{+}\nu_{e})$&$ \mathcal{B} r(B_{c}^{+}\to \eta_{c} \mu^{+}\nu_{\mu})$&$\mathcal{B} r(B_{c}^{+}\to \eta_{c} \tau^{+}\nu_{\tau})$&$ \mathcal{B} r(B_{c}^{+}\to J/\psi e^{+}\nu_{e})$&$ \mathcal{B} r(B_{c}^{+}\to J/\psi \mu^{+}\nu_{\mu})$&$ \mathcal{B} r(B_{c}^{+}\to J/\psi \tau^{+}\nu_{\tau}) $\\
  \hline
This work&$ 0.71^{+0.01+0.00+0.03}_{-0.01-0.00-0.03}$&$0.69^{+0.01+0.00+0.03}_{-0.01-0.00-0.03}$&$0.20^{+0.00+0.00+0.01}_{-0.00-0.01-0.01}$&$1.60^{+0.03+0.01+0.19}_{-0.03-0.19-0.04}$&$1.59^{+0.03+0.01+0.19}_{-0.03-0.20-0.04}$&$0.40^{+0.01+0.00+0.05}_{-0.01-0.04-0.02}$\\
\hline
\cite{C.F131,Qiao:132}&$ 2.1$&$2.1$&$0.64$&$6.7$&$6.7$&$0.52$\\
\cite{R.R}&$ 0.55$&$-$&$-$&$1.73$&$-$&$-$\\
\cite{Ebert10,Ebert101}&$ 0.42$&$-$&$-$&$1.23$&$-$&$-$\\
\cite{Ivanov06}&$ 0.81$&$-$&$0.22$&$2.07$&$-$&$0.49$\\
\cite{Huang07}&$ 1.64$&$-$&$0.49$&$2.37$&$-$&$0.65$\\
\cite{Wang09}&$ 0.67$&$0.67$&$0.19$&$1.49$&$1.49$&$0.37$\\
\cite{Hernndez06}&$ 0.48$&$0.48$&$0.17$&$1.54$&$1.54$&$0.41$\\
\cite{P00}&$ 0.15$&$-$&$-$&$1.5$&$-$&$-$\\
\cite{Issadykov:2017wlb}&$ 0.95$&$0.95$&$0.24$&$1.67$&$1.67$&$0.40$\\
\cite{Rui:2016opu}&$ 4.5$&$4.5$&$2.8$&$5.7$&$5.7$&$1.7$\\
\cite{Ivanov:2000aj}&$ 0.98$&$-$&$0.27$&$2.30$&$-$&$0.59$\\
\cite{Kiselev:2000pp,Kiselev:2002vz}&$ 0.75$&$0.75$&$0.23$&$1.9$&$1.9$&$0.48$\\
\cite{CHEN49}&$ 0.97$&$-$&$-$&$2.35$&$-$&$-$\\
\cite{Anisimov:1998uk,Anisimov:1998uk1}&$ 0.59$&$0.59$&$0.20$&$1.20$&$1.20$&$0.34$\\
\cite{Wang:2012lrc}&$ 0.44$&$0.44$&$0.14$&$1.01$&$1.01$&$0.29$\\
\cite{Faustov:2022ybm}&$ 0.42$&$0.42$&$0.16$&$1.31$&$1.30$&$0.37$\\
\hline\hline
\end{tabular}\label{tab1q}
}
\end{center}
\end{table}
From Tab. \ref{tab1q}, one can find that the value of the ratio $R_{J / \psi}$
\be
R_{J / \psi}=\frac{\mathcal{B} r\left(B_{c}^{+} \rightarrow J / \psi \tau^{+} \nu_{\tau}\right)}{\mathcal{B} r\left(B_{c}^{+} \rightarrow J / \psi \mu^{+} \nu_{\mu}\right)}=0.25\pm0.04,
\en
which is fall in the range of $0.24\le R_{J / \psi}\le 0.28$ predicted by most of other SM approaches, and is also consistent with the model-independent constraint $0.20\le R_{J / \psi}\le0.39$ \cite{Cohen:2018dgz}. Our prediction is smaller than the measured one shown in Eq. (\ref{rjpsi}), but still agrees with it within 2$\sigma$ errors. As a complementary check, we also calculate the value of $R_{\eta_{c}}$,
\be
R_{\eta_{c}}=\frac{\mathcal{B} r\left(B_{c}^{+} \rightarrow \eta_{c} \tau^{+} \nu_{\tau}\right)}{\mathcal{B} r\left(B_{c}^{+} \rightarrow \eta_{c} \mu^{+} \nu_{\mu}\right)}=0.29\pm0.02,
\en
 which is still consistent with most of other theoretical results lying in range of $0.25\le R_{\eta_{c}}\le 0.35$ \cite{Berns:2018vpl}, and agrees well with the model-independent prediction $0.29\pm0.05$\cite{Lamm:2018xmc}.

\begin{table}[H]
\caption{Branching ratios of the semileptonic $B_{c}$ decays to the radially excited charmonium states.}
\setlength{\tabcolsep}{0.3mm}{
\begin{center}
\begin{tabular}{c|c|cccccc}
\hline\hline
 Reference&This work &\cite{R.R}\;\;&\cite{Ebert10,Ebert101}&\;\;\cite{M08} &\;\;\cite{P00}&\;\;\cite{I}&\;\;\cite{Rui:2016opu}\\
 \hline
$10^{-3}\times \mathcal{B} r(B_{c}^{+}\to \eta_{c}(2S) e^{+}\nu_{e})$&$0.91^{+0.02+0.00+0.87}_{-0.02-0.00-0.69}$&$0.7$&$0.32$&$1.1$&$0.2$&$0.46$&$7.7$\\
$10^{-3}\times \mathcal{B} r(B_{c}^{+}\to \eta_{c}(2S) \mu^{+}\nu_{\mu})$&$0.91^{+0.02+0.00+0.87}_{-0.02-0.00-0.69}$&$-$&$-$&$-$&$-$&$-$&$-$\\
$10^{-5} \times \mathcal{B} r(B_{c}^{+}\to \eta_{c}(2S) \tau^{+}\nu_{\tau})$&$8.22^{+0.14+0.02+7.37}_{-0.14-0.02-6.52}$&$-$&$-$&$8.1$&$-$&$1.3$&$53$\\
\hline
$10^{-3}\times \mathcal{B} r(B_{c}^{+}\to \eta_{c}(3S) e^{+}\nu_{e})$&$0.33^{+0.01+0.00+0.11}_{-0.01-0.00-0.11}$&$-$&$0.0055$&$0.19$&$-$&$-$&$1.4$\\
$10^{-3}\times \mathcal{B} r(B_{c}^{+}\to \eta_{c}(3S) \mu^{+}\nu_{\mu})$&$0.33^{+0.01+0.00+0.11}_{-0.01-0.00-0.11}$&$-$&$-$&$-$&$-$&$-$&$-$\\
$10^{-5} \times \mathcal{B} r(B_{c}^{+}\to \eta_{c}(3S) \tau^{+}\nu_{\tau})$&$0.39^{+0.01+0.00+0.13}_{-0.01-0.00-0.12}$&$-$&$0.0005$&$0.57$&$-$&$-$&$0.19$\\
\hline\hline
$10^{-3} \times \mathcal{B} r(B_{c}^{+}\to \psi(2S) e^{+}\nu_{e})$&$2.34^{+0.04+0.04+0.09}_{-0.04-0.08-0.02}$&$1$&$0.31$&$-$&$1.2$&$2.1$&$12$\\
$10^{-3} \times \mathcal{B} r(B_{c}^{+}\to \psi(2S) \mu^{+}\nu_{\mu})$&$2.32^{+0.04+0.04+0.09}_{-0.04-0.08-0.02}$&$-$&$-$&$-$&$-$&$-$&$-$\\
$10^{-5} \times \mathcal{B} r(B_{c}^{+}\to \psi(2S) \tau^{+}\nu_{\tau})$&$15.89^{+0.28+0.06+0.09}_{-0.28-0.01-0.20}$&$-$&$-$&$-$&$-$&$15$&$84$\\
\hline
$10^{-3} \times \mathcal{B} r(B_{c}^{+}\to \psi(3S) e^{+}\nu_{e})$&$0.73^{+0.01+0.00+0.05}_{-0.01-0.00-0.05}$&$-$&$0.0057$&$-$&$-$&$-$&$0.36$\\
$10^{-3} \times \mathcal{B} r(B_{c}^{+}\to \psi(3S) \mu^{+}\nu_{\mu})$&$0.72^{+0.01+0.00+0.05}_{-0.01-0.00-0.05}$&$-$&$-$&$-$&$-$&$-$&$-$\\
$10^{-5} \times \mathcal{B} r(B_{c}^{+}\to \psi(3S) \tau^{+}\nu_{\tau})$&$0.58^{+0.01+0.00+0.03}_{-0.01-0.01-0.02}$&$-$&$0.0036$&$-$&$-$&$-$&$0.038$\\
\hline\hline
\end{tabular}\label{tab2}
\end{center}}
\end{table}
 The branching ratios of the semileptonic $B_{c}$ decays to the radially excited  charmonium states are shown in Tab. \ref{tab2}, together with the results obtained in other approaches for comparision. The uncertainties are the same with those shown in Tab. \ref{tab1q}. It is noticed that the large errors in the decays $B_{c}^{+}\to \eta_{c}(2S) \ell^{+}\nu_{\ell}$ are induced by the decay constant $f_{\eta_c(2S)}=(243^{+79}_{-111})$ MeV \cite{Zhang:2023ypl}. For the decays $B_c^+\to \eta_c(2S)\ell^+\nu_{\ell}, \psi(2S)\ell^+\nu_{\ell}$, their branching ratios  are much smaller than the PQCD calcualations \cite{Rui:2016opu}.   Our predictions are consistent with the results given by the BS equation \cite{R.R}, the light-cone QCD sum rules \cite{M08} and the ISGW2 quark model \cite{I}. Although the reults for the semileptonic $B_c$ decays to the ground state charmonia given by the relativistic quark model \cite{Ebert10,Ebert101} are appropriate, those for the semileptonic $B_c$ decays to the radially excited charmonium states given by such approach seem to be too small. The branching ratios of these semileptonic decays show a clear hierarchical relationship,
\be
\mathcal{B} r(B_{c} \rightarrow \eta_c(3S)\ell\nu_{\ell})&<& \mathcal{B} r(B_{c} \rightarrow \eta_c(2S)\ell\nu_{\ell})< \mathcal{B} r(B_{c} \rightarrow \eta_c(1S)\ell\nu_{\ell}), \\
\mathcal{B} r(B_{c} \rightarrow \psi(3S)\ell\nu_{\ell})&<& \mathcal{B} r(B_{c} \rightarrow \psi(2S)\ell\nu_{\ell})< \mathcal{B} r(B_{c} \rightarrow \psi(1S)\ell\nu_{\ell}).
\en
The main reason is that the relationships $m_{\eta_c(1S)}<m_{\eta_c(2S)}<m_{\eta_c(3S)}$ and $m_{\psi(1S)}<m_{\psi(2S)}<m_{\psi(3S)}$ lead to the phase spaces of the final states decrease with the increasing of the radial quantum number $n$.
 %At the same time, In Tab. \ref{tab1q}, we can see the decay process $B_{c}^{+}\to J/\psi \ell^{+}\nu_{\ell}$. The proportion of longitudinal polarization and %transverse polarization is almost half each, while the proportion of transverse polarization in the decay process $B_{c}^{+}\to J/\psi \tau^{+}\nu_{\tau}$ is %relatively large.  In order to reduce the theoretical errors,
%In Tab. \ref{tab2}.

 Similar with $R_{J/\Psi}$, it is useful to define the ratios of the branching ratios for the $B_c$ decays to the radially excited charmonia,
  where the uncertainties induced by the model calculations and the CKM matrix elements can be cancelled,
 \be
 R_{\psi(nS)}=\frac{\mathcal{B} r(B^+_{c} \to \psi(nS)\tau^+ \nu_{\tau})}{\mathcal{B} r(B^+_{c} \rightarrow \psi(nS)\mu^+\nu_{\mu})},
 \;\; R_{\eta_c(nS)}=\frac{\mathcal{B} r(B^+_{c} \to \eta_c(nS)\tau^+ \nu_{\tau})}{\mathcal{B} r(B^+_{c} \rightarrow \eta_c(nS)\mu^+\nu_{\mu})},
 \en
 where $n=2, 3$. Our predictions $R_X$ with $X=\eta_c(1S,2S,3S), \psi(1S,2S,3S)$ are collected in Tab. \ref{tabn}, together with the results obtained in other approaches for comparison. Obviously, it shows a clear hierarchical relationship,
\be
R_{\psi(3S)}< R_{\psi(2S)}< R_{J/\Psi}, \;\;\; R_{\eta_c(3S)}< R_{\eta_c(2S)}< R_{\eta_c(1S)}.
\en
From Tab. \ref{tabn}, one can find that our predictions for these $R$ values are comparable to other SM calculations. If the departures of the SM predictions for $R_{\psi(2S, 3S)}$ and $R_{\eta_c(1S, 2S, 3S)}$ from the experimental data can be detected, it will further highlight the puzzle
in flavor physics and the failure of lepton flavour universality hinted in the $R_{J/\Psi}$ measurement \cite{LHCb:2017vlu}.  As our predictions for
the branching ratios of these semileptonic $B_c$ decays to the radially excited charmonia are larger than $10^{-6}$, which can be measured in the future High-luminosity LHC (HL-LHC) and High-energy LHC (HE-LHC) experiments \cite{Cerri:2018ypt}. Then, the semileptonic $B_c$ decays to the ground and excited charmonia will provide a more complete research area.

\begin{table}[H]
\caption{The values of ratios $R_X$ with $X=\eta_c(1S,2S,3S), \psi(1S,2S,3S)$.}
\setlength{\tabcolsep}{0.3mm}{
\begin{center}
\begin{tabular}{|c|c|c|c|c|c|c|c|c|c|c|}
\hline
 Ratio&This work &\cite{Nayak:2022gdo}&\cite{Ivanov06}&\cite{ZZT}&\cite{Wang:2012lrc}&\cite{Ebert10,Ebert101}&\cite{JFSD}&\cite{Issadykov:2017wlb} &\cite{M08}&\cite{I}\\
 \hline
  $R_{\eta_{c}}$&$0.29\pm0.02$&$-$&$0.27$&$-$&$0.31$&$-$&$-$&$0.25$&$-$&$-$\\
 \hline
 $R_{\eta_{c}(2S)}$&$0.09\pm0.12$&$0.14$&$-$&$0.054$&$-$&$-$&$0.069$&$-$&$0.74$&$0.028$\\
 \hline
 $R_{\eta_{c}(3S)}$&$0.012\pm0.006$&$0.021$&$-$&$0.010$&$-$&$0.0010$&$0.0014$&$-$&$0.03$&$-$\\
 \hline
 $R_{J/\psi}$&$0.25\pm0.04$&$-$&$0.24$&$-$&$0.29$&$-$&$-$&$0.24$&$-$&$-$\\
 \hline
 $R_{\psi(2S)}$&$0.068\pm0.003$&$0.085$&$-$&$0.51$&$-$&$-$&$0.070$&$-$&$-$&$0.071$\\
 \hline
 $R_{\psi(3S)}$&$0.0081\pm0.0007$&$0.078$&$-$&$0.0092$&$-$&$0.0063$&$0.0011$&$-$&$-$&$-$\\
 \hline
\end{tabular}\label{tabn}
\end{center}}
\end{table}

In order to investigate the impact of the lepton masses and provide more detailed physical picture for the semilepton $B_c$ decays beyond the branching ratio, we also define another two physical observations that can be measured by experiments: the longitudinal polarization fraction $f_{L}$ and the forward-backward asymmetry $A_{FB}$. These two physical quantities are sensitive to some kinds
 of new physics \cite{ptau2,prd072012,prd114022,prd036021,ptau1}, so their values are helpful to test the SM and different NP scenarios. Meanwhile,
 the calculations of these two quantities may provide new clues to understand the ratio $R_{J/\Psi}$ puzzle. We expect that these physical obervables can be measured by the future LHCb experiments and it is helpful to clarify dispution from the different theoretical approaches. From the numerical results listed in Tab. \ref{tab1xfl} and Tab. \ref{tab1x}, the following points can be found:

\begin{table}[H]
\caption{The longitudinal polarization fractions $f_{L}$ (in\%) for the decays $B_{c}^+\to \eta_{c}(1S,2S,3S)\ell^+\nu_{\ell}$, $B_{c}^+\to\psi(1S,2S,3S)\ell^+\nu_{\ell}$. The uncertainties are the same with those given in Table \ref{tab1q}.}
\begin{center}
\scalebox{0.8}{
\begin{tabular}{|c|c|c|c|}
\hline\hline
Decay modes&$B_{c}^{+}\to J/\psi e^{+}\nu_{e}$&$B_{c}^{+}\to J/\psi \mu^{+}\nu_{\mu}$&$B_{c}^{+}\to J/\psi \tau^{+}\nu_{\tau}$\\
 \hline\hline
This work&$50.9^{+0.9+2.9+5.5}_{-0.9-4.6-5.1}$&$50.9^{+0.9+3.0+5.7}_{-0.9-4.7-5.2}$&$43.8^{+0.8+3.6+5.6}_{-0.8-4.7-5.1}$\\
\cite{Rui:2016opu}&$33$&$33$&$39$\\
\cite{Faustov:2022ybm}&$44$&$44$&$40$\\
\cite{Wang:2008xt}&$51$&$51$&$45$\\
\hline\hline
Decay modes&$B_{c}^{+}\to \psi(2S) e^{+}\nu_{e}$&$B_{c}^{+}\to \psi(2S) \mu^{+}\nu_{\mu}$&$B_{c}^{+}\to \psi(2S) \tau^{+}\nu_{\tau}$\\
 \hline\hline
This work&$59.2^{+1.0+0.1+0.1}_{-1.0-0.1-0.2}$&$59.2^{+1.0+0.1+0.1}_{-1.0-0.1-0.2}$&$44.6^{+0.8+0.1+0.2}_{-0.8-0.3-0.3}$\\
\cite{Rui:2016opu}&$46$&$46$&$41$\\
\hline\hline
 Decay modes&$B_{c}^{+}\to \psi(3S) e^{+}\nu_{e}$&$B_{c}^{+}\to \psi(3S) \mu^{+}\nu_{\mu}$&$B_{c}^{+}\to \psi(3S) \tau^{+}\nu_{\tau}$\\
 \hline\hline
This work&$57.9^{+1.0+0.2+4.0}_{-1.0-0.1-4.1}$&$57.9^{+1.0+0.2+4.0}_{-1.0-0.1-4.1}$&$40.5^{+0.7+0.1+3.2}_{-0.7-0.2-3.1}$\\
\cite{Rui:2016opu}&$54$&$54$&$31$\\
\hline\hline
\end{tabular}\label{tab1xfl}}
\end{center}
\end{table}
(1) From Eqs.(\ref{eq:decaywidthlon})-(\ref{eq:fl}), we find that the longitudinal polarization fractions $f_{L}$ of the decays $B_{c}^{+}\to J/\psi \ell^{+}\nu_{\ell}$ decrease with the $m_{\ell}$ increasing although such trend is mild, that is
\be
 f_L(B_{c}^{+}\to J/\psi e^{+}\nu_{e})\approx f_L(B_{c}^{+}\to J/\psi \mu^{+}\nu_{\mu})>f_L(B_{c}^{+}\to J/\psi \tau^{+}\nu_{\tau}),
\en
which is support by our predictions shown in Tab. \ref{tab1xfl}. Certainly, this rule can also be  applied to the decays $B_{c}^{+}\to \psi(2S,3S) \ell^{+}\nu_{\ell}$.
\begin{table}[H]
\caption{Partial branching ratios and longitudinal polarization fraction $f_{L}$(in\%) for the decays $B_{c}^+\to \eta_{c}(1S,2S,3S)\ell^+\nu_{\ell}$, $B_{c}^+\to\psi(1S,2S,3S)\ell^+\nu_{\ell}$ in Region 1 and Region 2. }
\begin{center}
\scalebox{0.7}{
\begin{tabular}{|c|c|c|c|c|c|c|c|c|}
\hline\hline
Observables&Region 1&Region 2&Observables&Region 1&Region 2&Observables&Region 1&Region 2\\
\hline
$\mathcal{B} r(B_{c}^{+}\to J/\psi e^{+}\nu_{e})$&$0.71\times10^{-2}$&$0.89\times10^{-2}$&$\mathcal{B} r(B_{c}^{+}\to J/\psi \mu^{+}\nu_{\mu})$&$0.71\times10^{-2}$&$0.88\times10^{-2}$&$\mathcal{B} r(B_{c}^{+}\to J/\psi \tau^{+}\nu_{\tau})$&$0.13\times10^{-2}$&$0.27\times10^{-2}$\\
$f_{L}(B_{c}^{+}\to J/\psi e^{+}\nu_{e})$&$64.5$&$40.1$&$f_{L}(B_{c}^{+}\to J/\psi \mu^{+}\nu_{\mu})$&$64.5$&$40.1$&$f_{L}(B_{c}^{+}\to J/\psi \tau^{+}\nu_{\tau})$&$53.9$&$40.1$\\
\hline
$\mathcal{B} r(B_{c}^{+}\to \psi(2S) e^{+}\nu_{e})$&$1.33\times10^{-3}$&$1.01\times10^{-3}$&$\mathcal{B} r(B_{c}^{+}\to \psi(2S) \mu^{+}\nu_{\mu})$&$1.31\times10^{-3}$&$1.01\times10^{-3}$&$\mathcal{B} r(B_{c}^{+}\to \psi(2S) \tau^{+}\nu_{\tau})$&$5.33\times10^{-5}$&$10.55\times10^{-5}$\\
$f_{L}(B_{c}^{+}\to \psi(2S) e^{+}\nu_{e})$&$71.2$&$43.4$&$f_{L}(B_{c}^{+}\to \psi(2S) \mu^{+}\nu_{\mu})$&$71.2$&$43.5$&$f_{L}(B_{c}^{+}\to \psi(2S) \tau^{+}\nu_{\tau})$&$52.4$&$40.7$\\
\hline
$\mathcal{B} r(B_{c}^{+}\to \psi(3S) e^{+}\nu_{e})$&$0.40\times10^{-3}$&$0.33\times10^{-3}$&$\mathcal{B} r(B_{c}^{+}\to \psi(3S) \mu^{+}\nu_{\mu})$&$0.40\times10^{-3}$&$0.33\times10^{-3}$&$\mathcal{B} r(B_{c}^{+}\to \psi(3S) \tau^{+}\nu_{\tau})$&$0.16\times10^{-5}$&$0.42\times10^{-5}$\\
$f_{L}(B_{c}^{+}\to \psi(3S) e^{+}\nu_{e})$&$70.2$&$42.8$&$f_{L}(B_{c}^{+}\to \psi(3S) \mu^{+}\nu_{\mu})$&$70.2$&$42.9$&$f_{L}(B_{c}^{+}\to \psi(3S) \tau^{+}\nu_{\tau})$&$45.7$&$38.1$\\
\hline\hline
\end{tabular}\label{tablex3}}
\end{center}
\end{table}

In order to investigate the dependences of the polarizations on the different $q^2$, we calculate the longitudinal polarization fractions
by dividing the full energy region into two regions for each decay, which are listed in Tab. \ref{tablex3}, together with the partial branching ratios, where
 Region 1 is defined as $m_{\ell}^{2}<q^{2}<\frac{(m_{B_{c}}-m_{\psi(nS)})^{2}+m_{\ell}^{2}}{2} $ and Region 2 is $\frac{(m_{B_{c}}-m_{\psi(nS)})^{2}+m_{\ell}^{2}}{2} <q^{2}<(m_{B_{c}}-m_{\psi(nS)})^{2}$ with $n=1,2,3$. From Tab. \ref{tablex3}, one can find that in Region 1 the longitudinal polarizations dominate the branching ratios for the decays with $e$ and $\mu$ involved, while the longitudinal and transverse polarizations for the decay channels involving $\tau$ are comparable in Region 1. It is interesting that for all the considered decays the transverse polarizations are dominant in Region 2. These results can be tested by the future LHCb experiments.
 \begin{table}[H]
\caption{The forward-backward asymmetries $A_{FB}$ for the decays $B_{c}^+\to \eta_{c}(1S,2S,3S)\ell^+\nu_{\ell}$ and $B_{c}^+\to\psi(1S,2S,3S)\ell^+\nu_{\ell}$.}
\begin{center}
\scalebox{0.85}{
\begin{tabular}{|c|c|c|c|}
\hline\hline
Decay modes&$B_{c}^{+}\to \eta_{c} e^{+}\nu_{e}$&$B_{c}^{+}\to \eta_{c} \mu^{+}\nu_{\mu}$&$B_{c}^{+}\to \eta_{c} \tau^{+}\nu_{\tau}$\\
 \hline\hline
This work&$(3.93^{+0.07+0.00+0.14}_{-0.07-0.00-0.01})\times 10^{-7}$&$0.015^{+0.000+0.000+0.001}_{-0.000-0.000-0.000}$&$0.36^{+0.01+0.01+0.02}_{-0.01-0.01-0.01}$\\
\cite{Faustov:2022ybm}&$-8.6\times 10^{-7}$&$-0.012$&$-0.35$\\
\cite{Nayak:2021djn}&$-2.049\times 10^{-7}$&$-$&$0.357$\\
\cite{Huang:2018nnq}&$-$&$-$&$0.364$\\
\hline\hline
Decay modes&$B_{c}^{+}\to \eta_{c}(2S) e^{+}\nu_{e}$&$B_{c}^{+}\to \eta_{c}(2S) \mu^{+}\nu_{\mu}$&$B_{c}^{+}\to \eta_{c}(2S) \tau^{+}\nu_{\tau}$\\
 \hline\hline
This work&$(6.96^{+0.12+0.00+5.69}_{-0.12-0.00-5.17})\times10^{-7}$&$0.023^{+0.000+0.000+0.019}_{-0.000-0.000-0.017}$&$0.30^{+0.01+0.00+0.28}_{-0.01-0.00-0.24}$\\
\cite{Nayak:2022gdo}&$5.75\times10^{-7}$&$-$&$0.384$\\
\hline\hline
 Decay modes&$B_{c}^{+}\to \eta_{c}(3S) e^{+}\nu_{e}$&$B_{c}^{+}\to \eta_{c}(3S) \mu^{+}\nu_{\mu}$&$B_{c}^{+}\to \eta_{c}(3S) \tau^{+}\nu_{\tau}$\\
 \hline\hline
This work&$(9.00^{+0.16+0.00+2.77}_{-0.16-0.00-2.88})\times 10^{-7}$&$0.028^{+0.000+0.000+0.009}_{-0.000-0.000-0.009}$&$0.33^{+0.01+0.00+0.14}_{-0.01-0.00-0.12}$\\
\cite{Nayak:2022gdo}&$10.5\times 10^{-7}$&$-$&$0.367$\\
\hline\hline
Decay modes&$B_{c}^{+}\to J/\psi e^{+}\nu_{e}$&$B_{c}^{+}\to J/\psi \mu^{+}\nu_{\mu}$&$B_{c}^{+}\to J/\psi \tau^{+}\nu_{\tau}$\\
 \hline\hline
This work&$-0.21^{+0.00+0.02+0.02}_{-0.00-0.02-0.02}$&$-0.21^{+0.00+0.02+0.02}_{-0.00-0.02-0.02}$&$-0.14^{+0.00+0.01+0.02}_{-0.00-0.01-0.02}$\\
\cite{Faustov:2022ybm}&$-0.19$&$-0.19$&$-0.23$\\
\cite{Nayak:2021djn}&$0.18$&$-$&$-0.255$\\
\cite{Huang:2018nnq}&$-$&$-$&$-0.042$\\
\hline\hline
Decay modes&$B_{c}^{+}\to \psi(2S) e^{+}\nu_{e}$&$B_{c}^{+}\to \psi(2S) \mu^{+}\nu_{\mu}$&$B_{c}^{+}\to \psi(2S) \tau^{+}\nu_{\tau}$\\
 \hline\hline
This work&$-0.16^{+0.00+0.00+0.00}_{-0.00-0.00-0.01}$&$-0.16^{+0.00+0.00+0.00}_{-0.00-0.00-0.01}$&$-0.093^{+0.002+0.000+0.014}_{-0.002-0.000-0.003}$\\
\cite{Nayak:2022gdo}&$-0.246$&$-$&$-0.214$\\
\hline\hline
Decay modes&$B_{c}^{+}\to \psi(3S) e^{+}\nu_{e}$&$B_{c}^{+}\to \psi(3S) \mu^{+}\nu_{\mu}$&$B_{c}^{+}\to \psi(3S) \tau^{+}\nu_{\tau}$\\
 \hline\hline
This work&$-0.14^{+0.00+0.00+0.01}_{-0.00-0.00-0.01}$&$-0.14^{+0.00+0.00+0.01}_{-0.00-0.00-0.01}$&$-0.048^{+0.001+0.009+0.014}_{-0.001-0.008-0.003}$\\
\cite{Nayak:2022gdo}&$-0.155$&$-$&$-0.144$\\
\hline\hline
\end{tabular}\label{tab1x}}
\end{center}
\end{table}
(2)From Tab. \ref{tab1x}, we find that the
 ratios of the forward-backward asymmetries $A_{FB}$ between the different semileptonic decays associated with the $B_{c}\rightarrow \eta_{c}(nS)$ transitions have the following rules: $ A_{FB}^{\mu}/A_{FB}^{e}=(3\sim4)\times10^{4}$ and $ A_{FB}^{\tau}/A_{FB}^{e} > 1\times10^{5}$. It is because that $A_{FB}$
 for $B_c\to \eta_c(nS)$ transitions are proportional to the square of the lepton mass shown in Eqs.(\ref{eq:AFB})-(\ref{eq:btheta1}). Undoubtedly, the effect of lepton mass can be well
 checked in such decay modes with the pseudoscalar mesons involved in the final states. As to the $B_{c}\to \psi(nS)$ transitions, the values of the forward-backward asymmetries $A_{FB}^{\mu}$ are almost equal to those of $A_{FB}^{e}$, while the values of $A_{FB}^{\tau}$ are smaller than those of $A_{FB}^{e(\mu)}$. It is noticed that dominant contribution to the $A_{FB}$ for the $B_c\to \psi(nS)$ transitions arises from the term proportional to $(H^{2}_{+}-H^{2}_{-})$ in Eq. (\ref{eq:btheta2}).

\begin{figure}[H]
\vspace{0.32cm}
  \centering
  \subfigure[]{\includegraphics[width=0.30\textwidth]{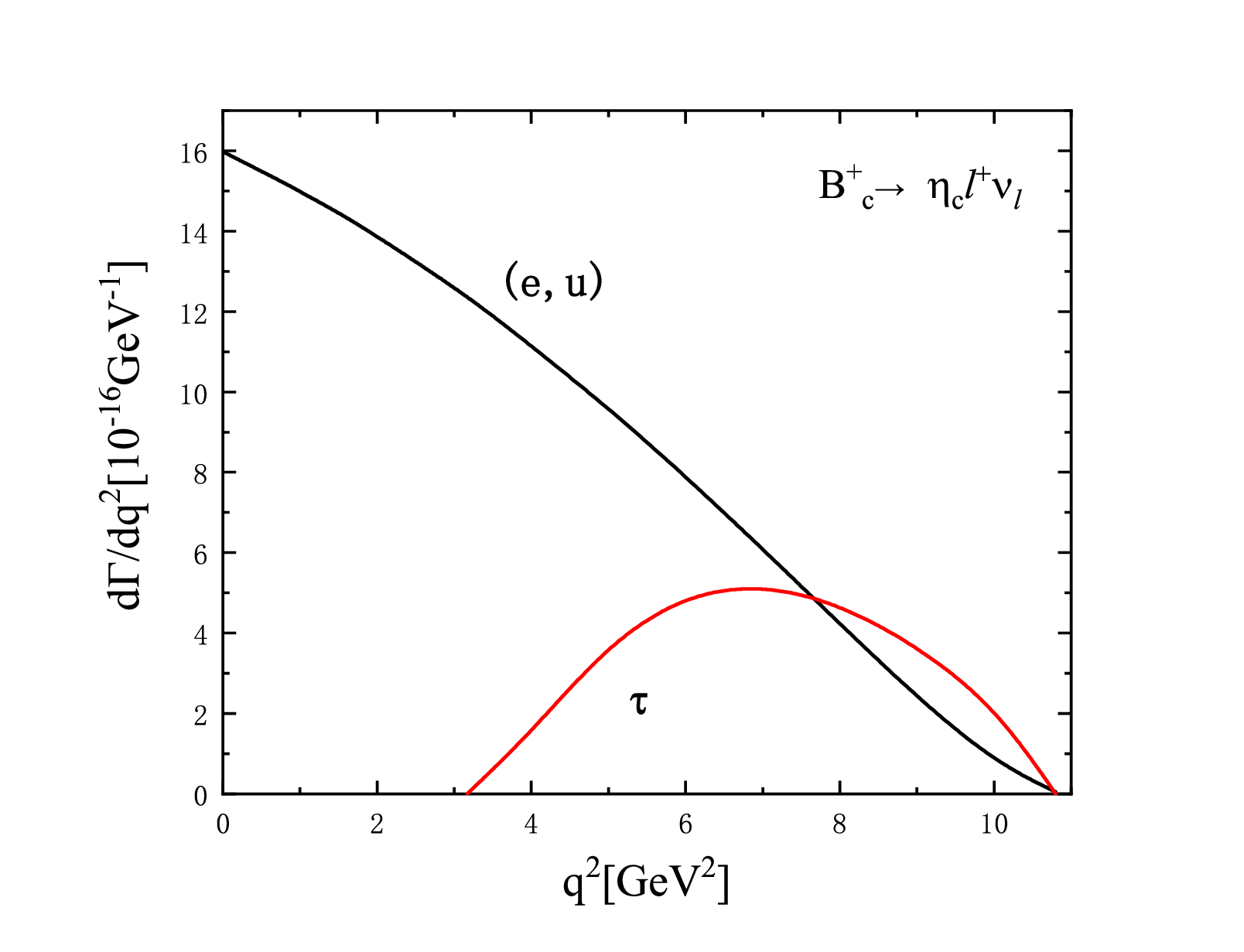}\quad}
  \subfigure[]{\includegraphics[width=0.30\textwidth]{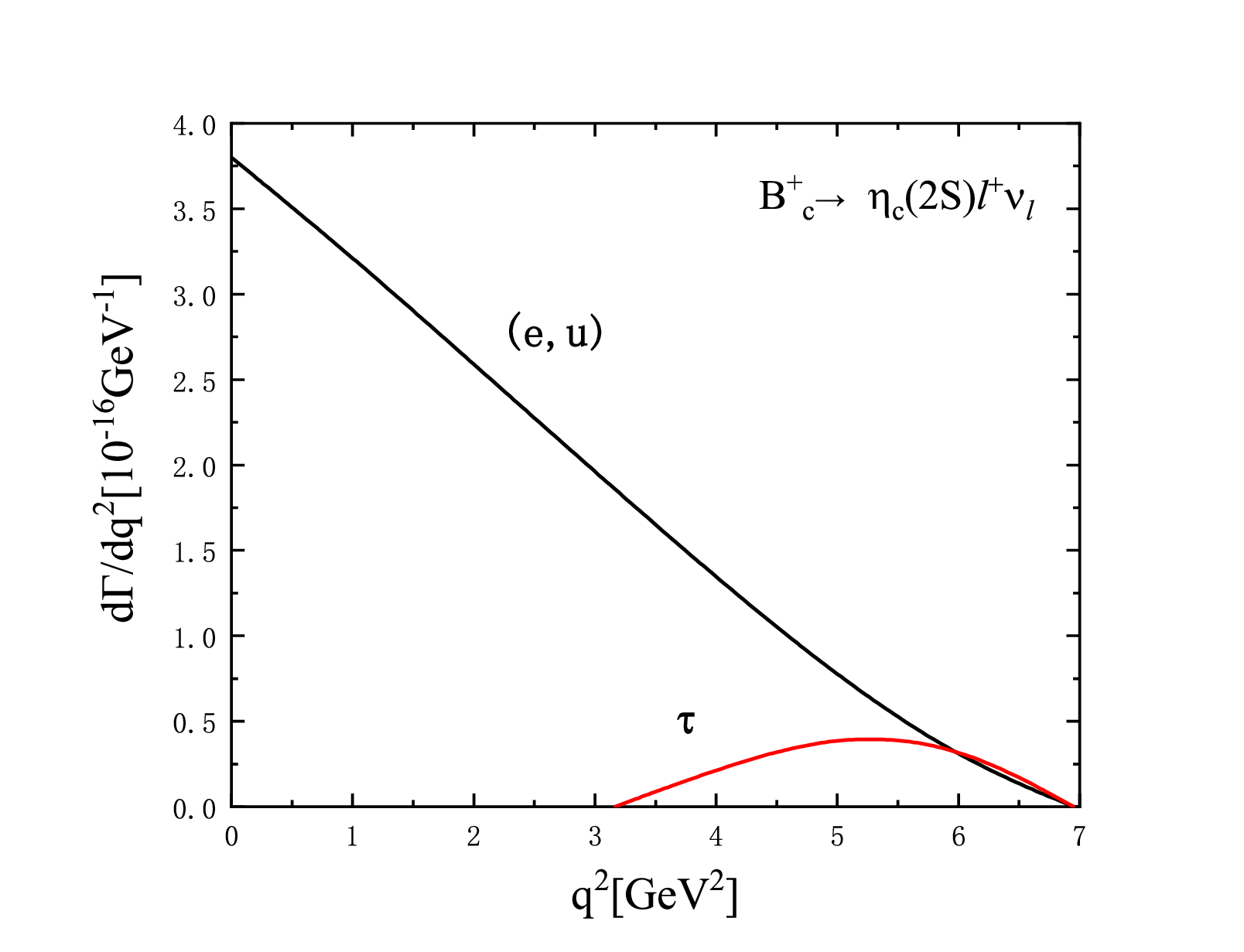}\quad}
  \subfigure[]{\includegraphics[width=0.30\textwidth]{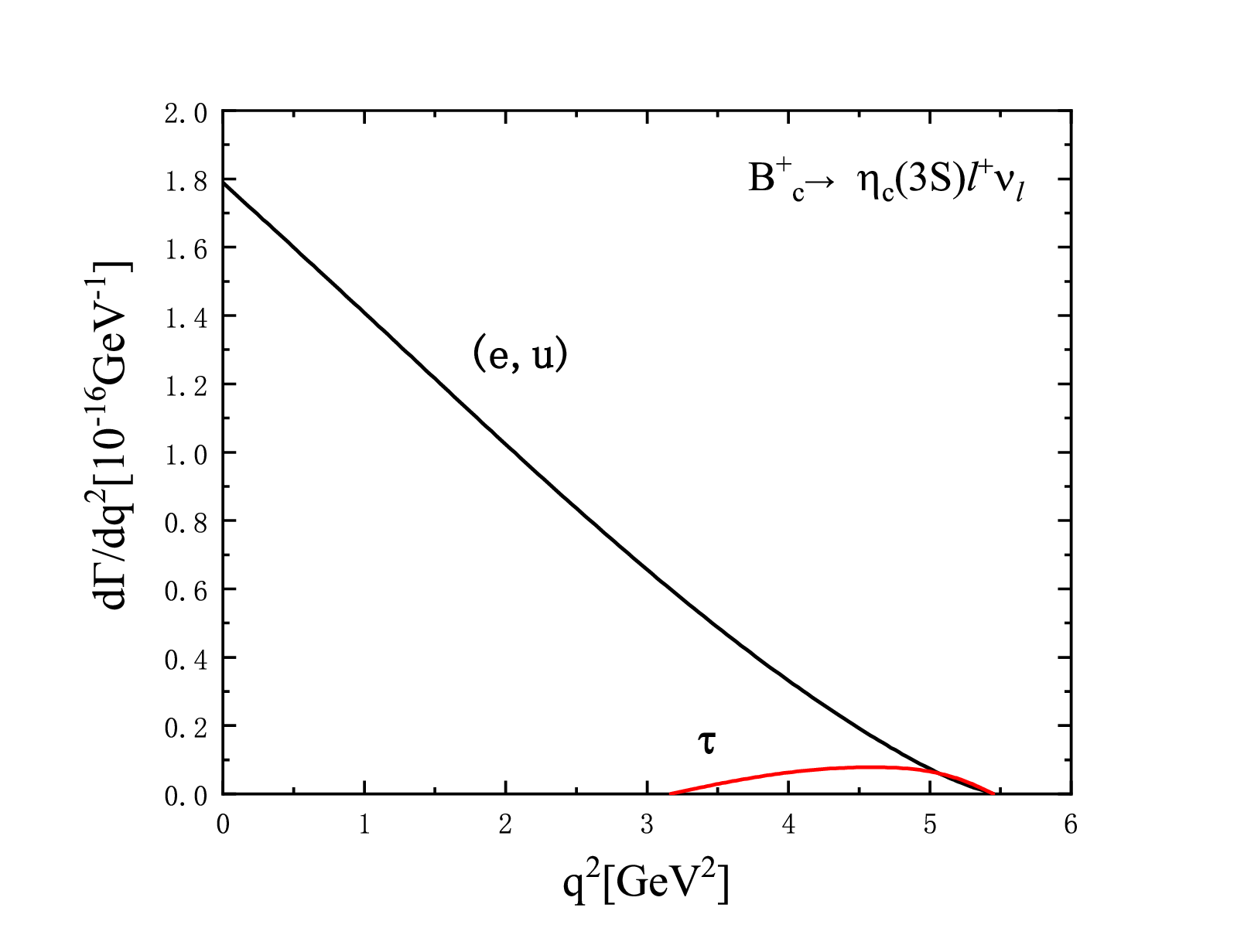}}\\
  \subfigure[]{\includegraphics[width=0.30\textwidth]{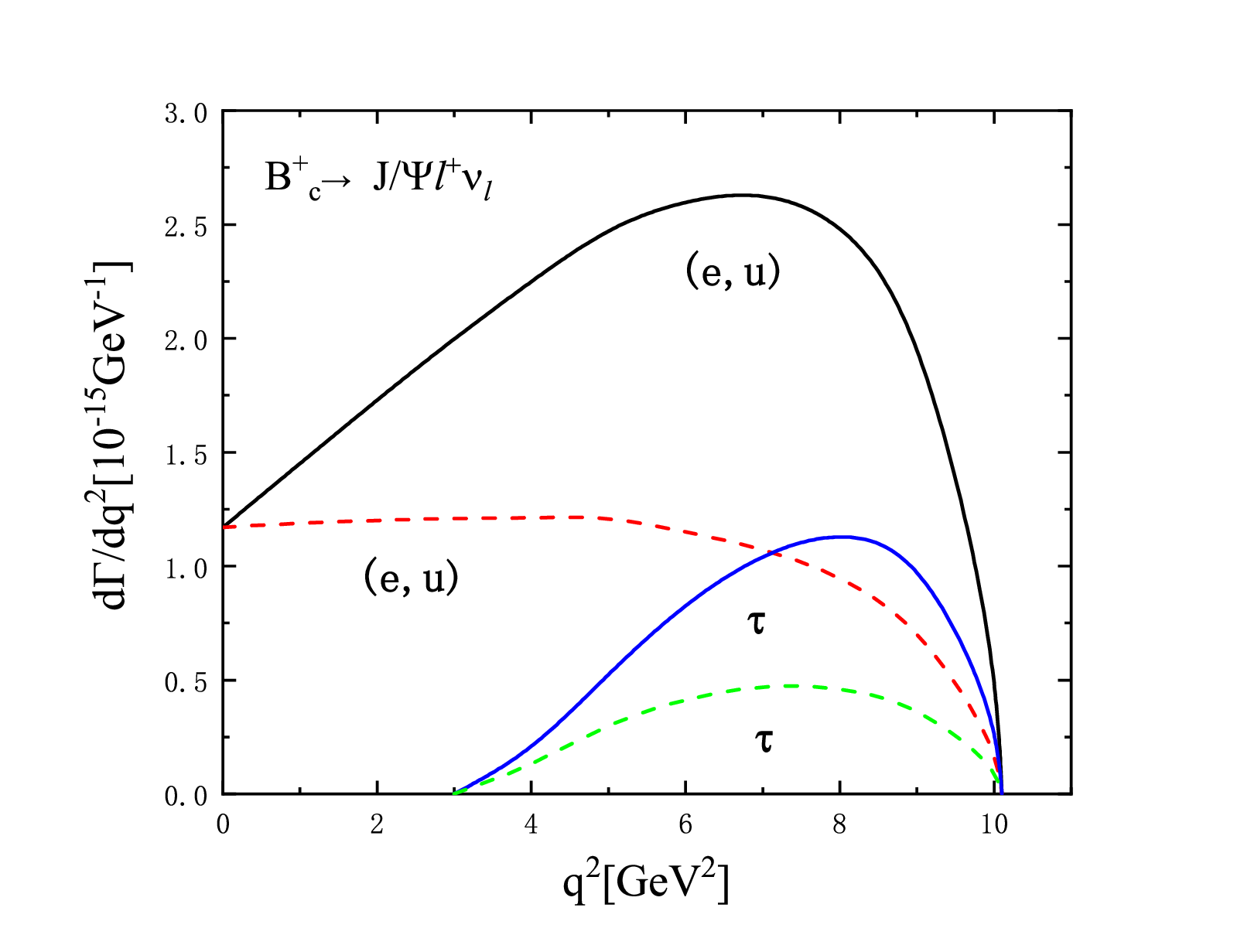}\quad}
  \subfigure[]{\includegraphics[width=0.30\textwidth]{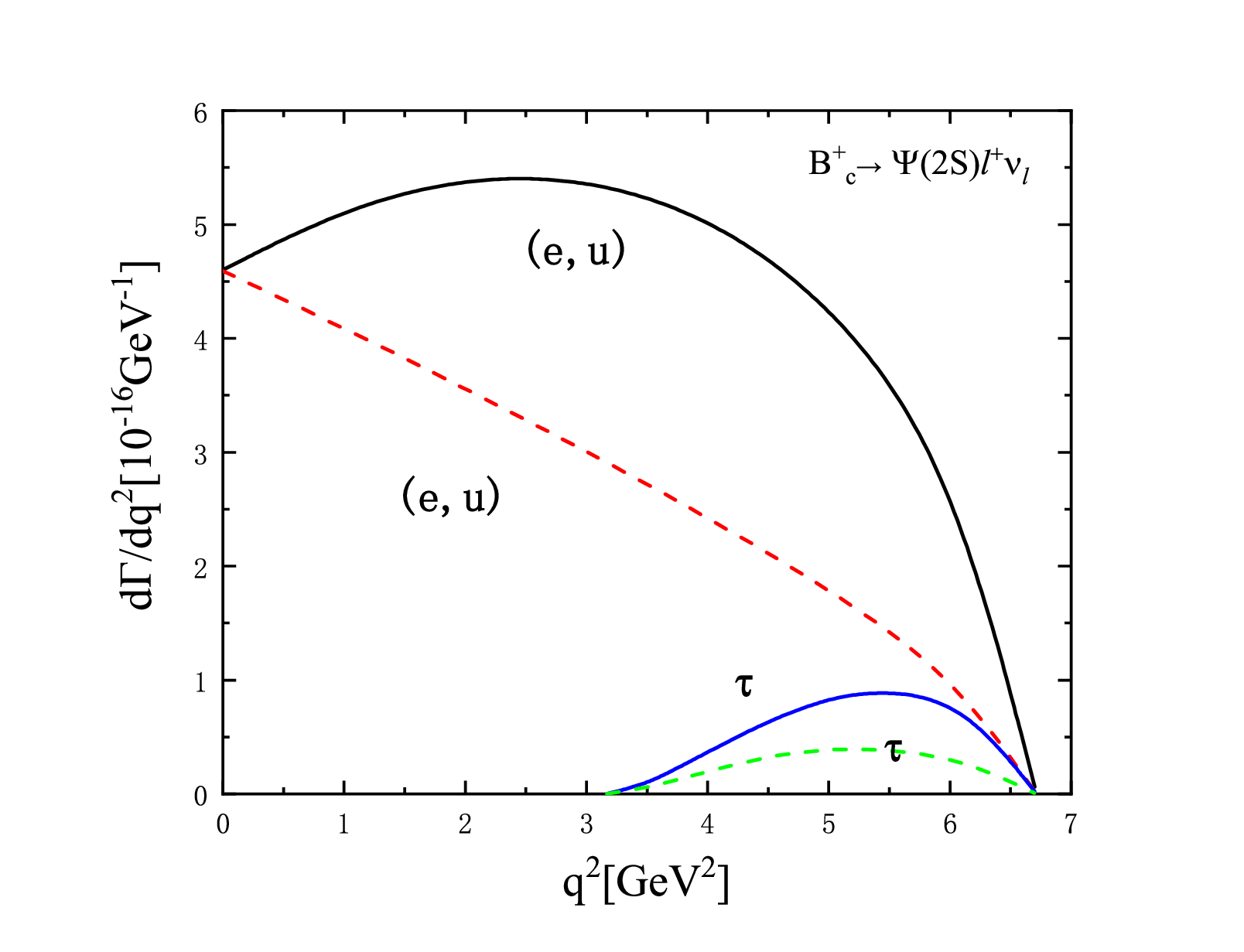}\quad}
  \subfigure[]{\includegraphics[width=0.30\textwidth]{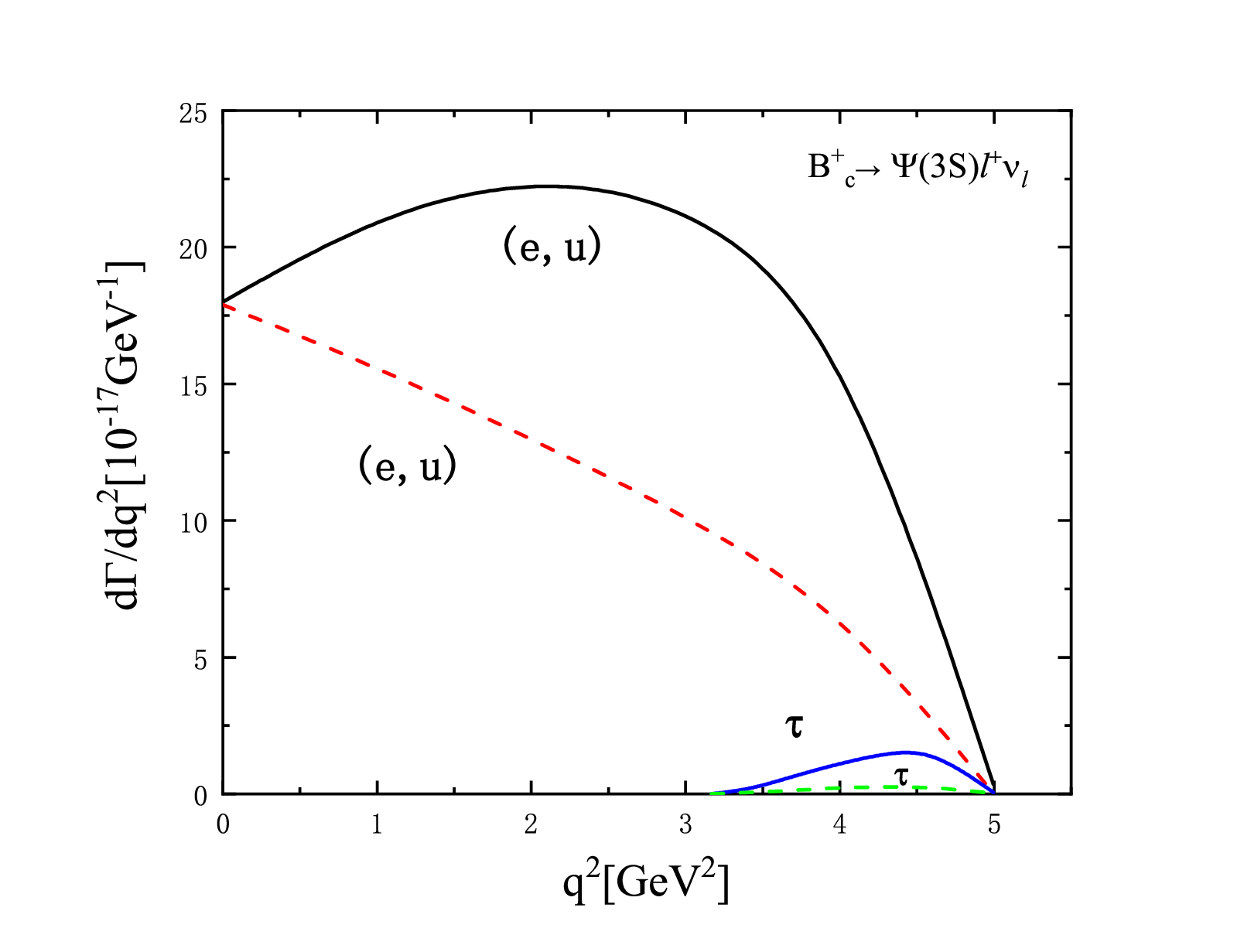}}
  \caption{The $q^{2}$ dependences of the differential decay rates $d\Gamma/dq^{2}$ (solid lines) and $d\Gamma_{L}/dq^{2}$ (dashed lines) with $L$ representing the contribution from the longitudinal polarization.}\label{fig:T22}
\end{figure}

\begin{figure}[H]
\vspace{0.32cm}
  \centering
  \subfigure[]{\includegraphics[width=0.30\textwidth]{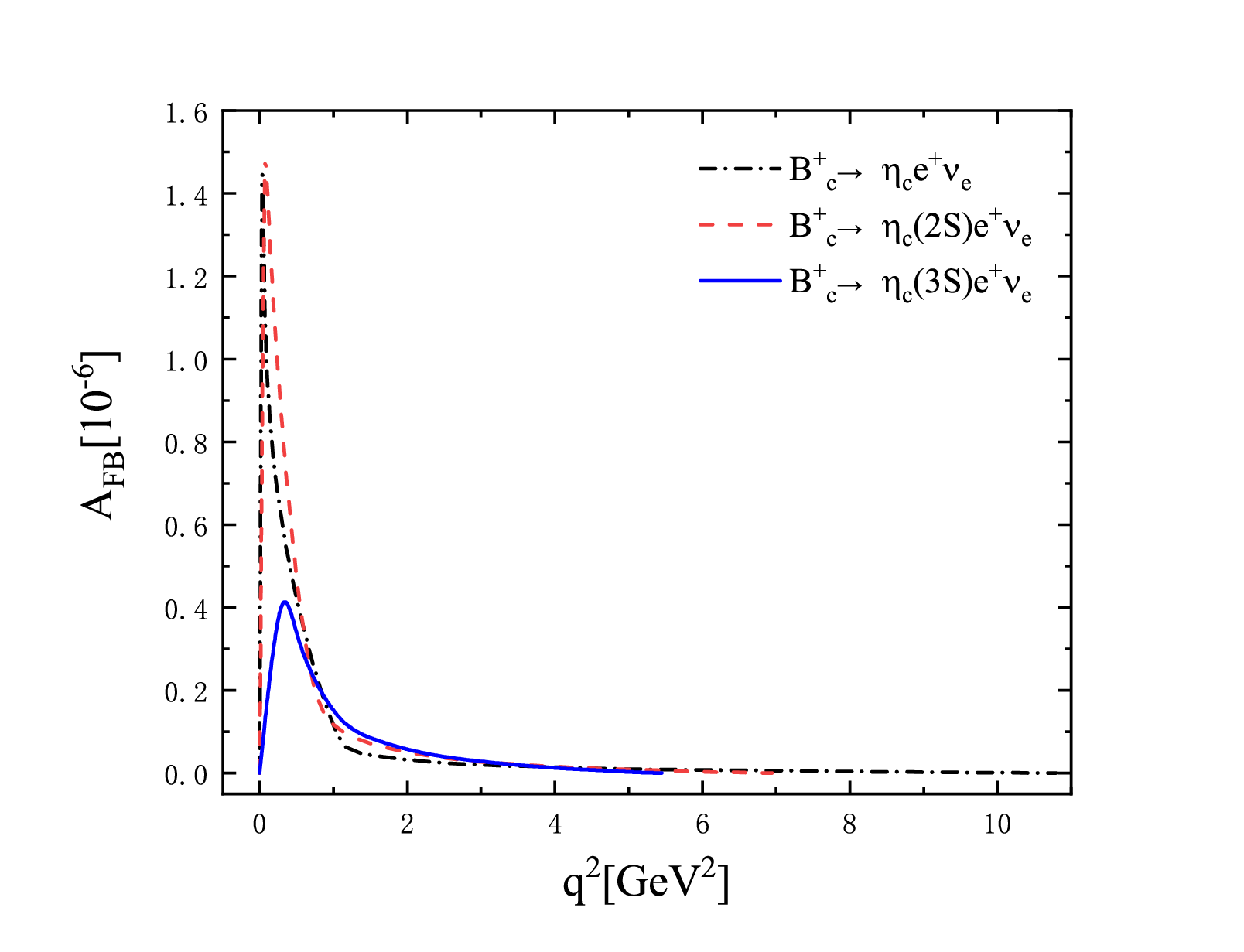}\quad}
  \subfigure[]{\includegraphics[width=0.30\textwidth]{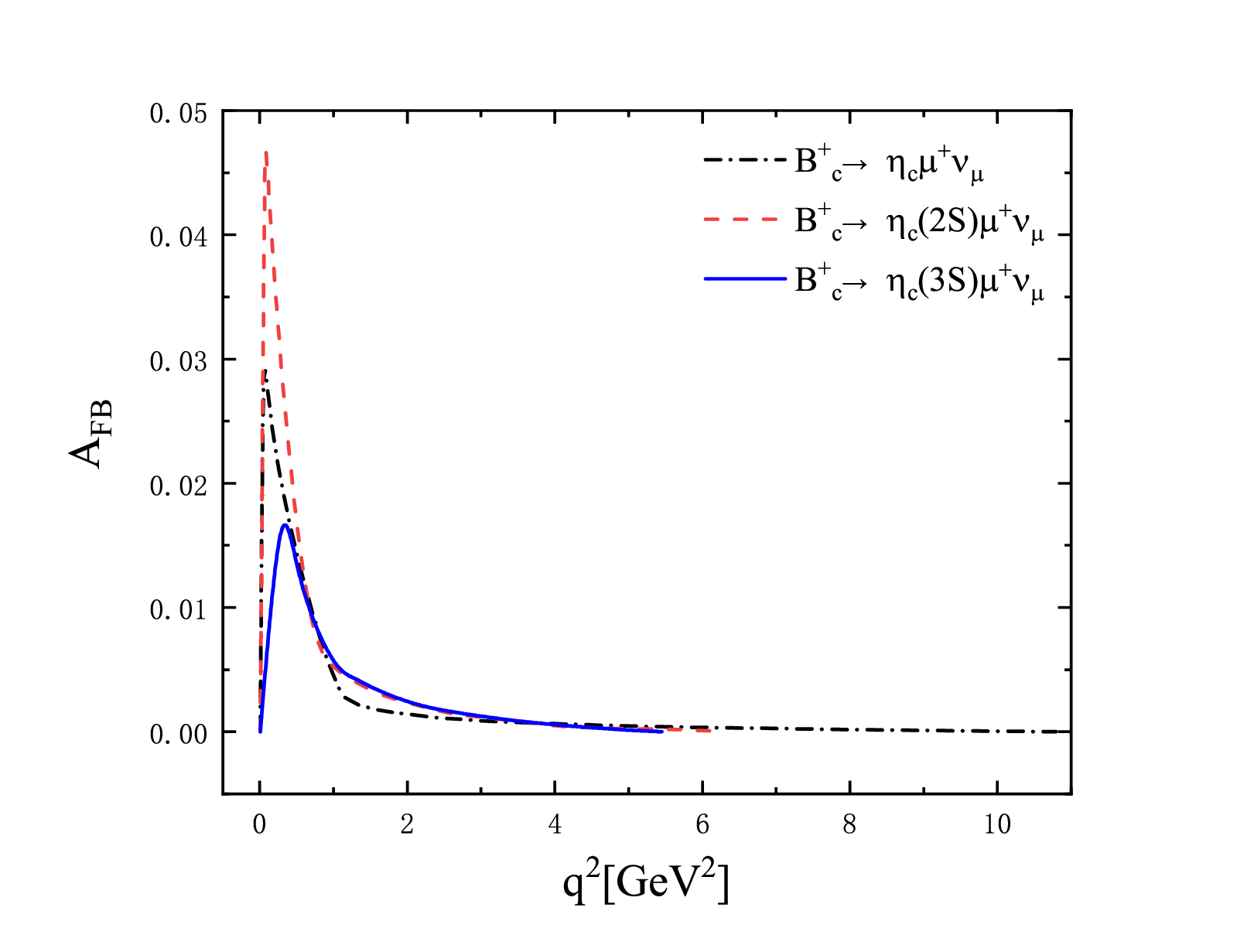}\quad}
  \subfigure[]{\includegraphics[width=0.30\textwidth]{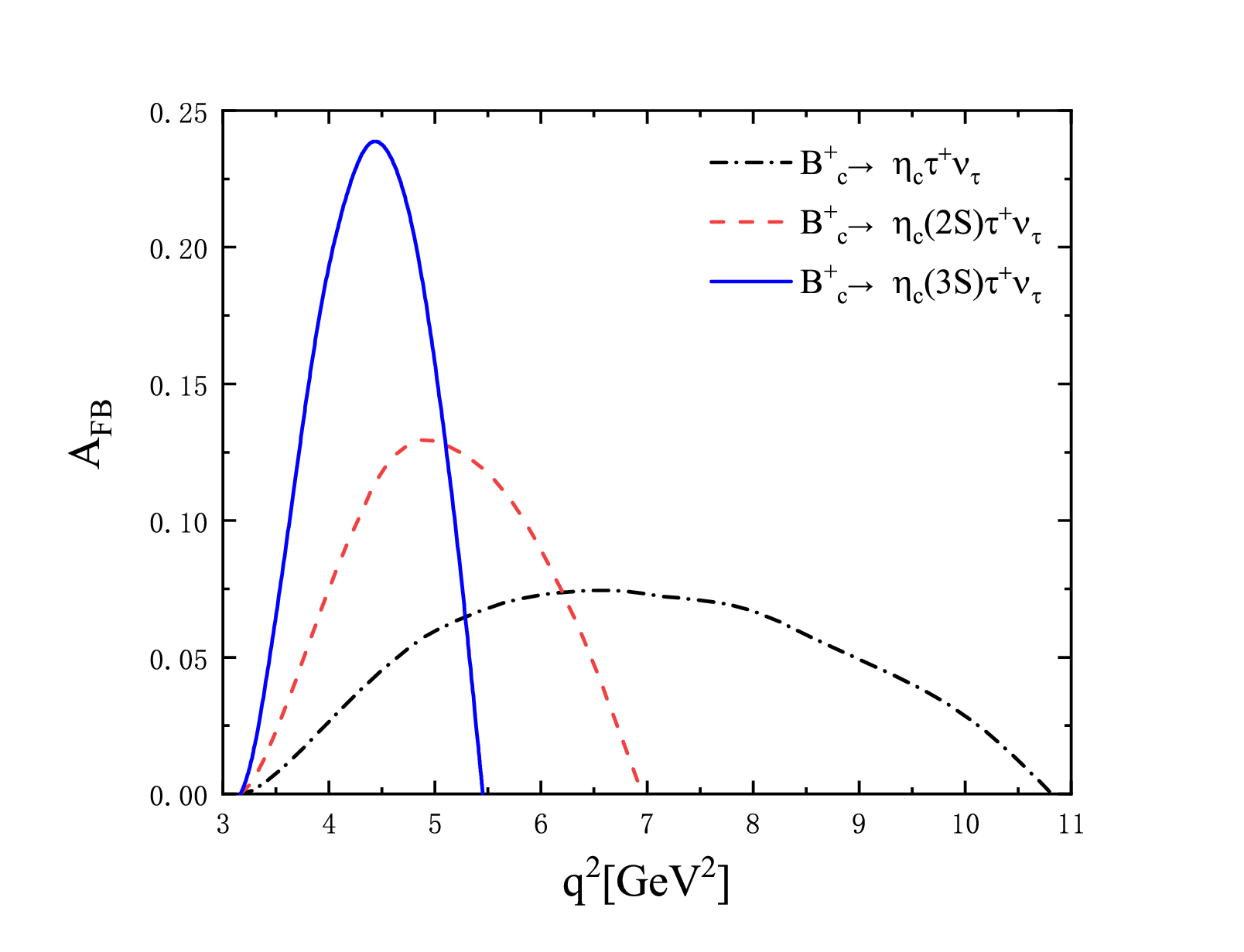}}\\
  \subfigure[]{\includegraphics[width=0.30\textwidth]{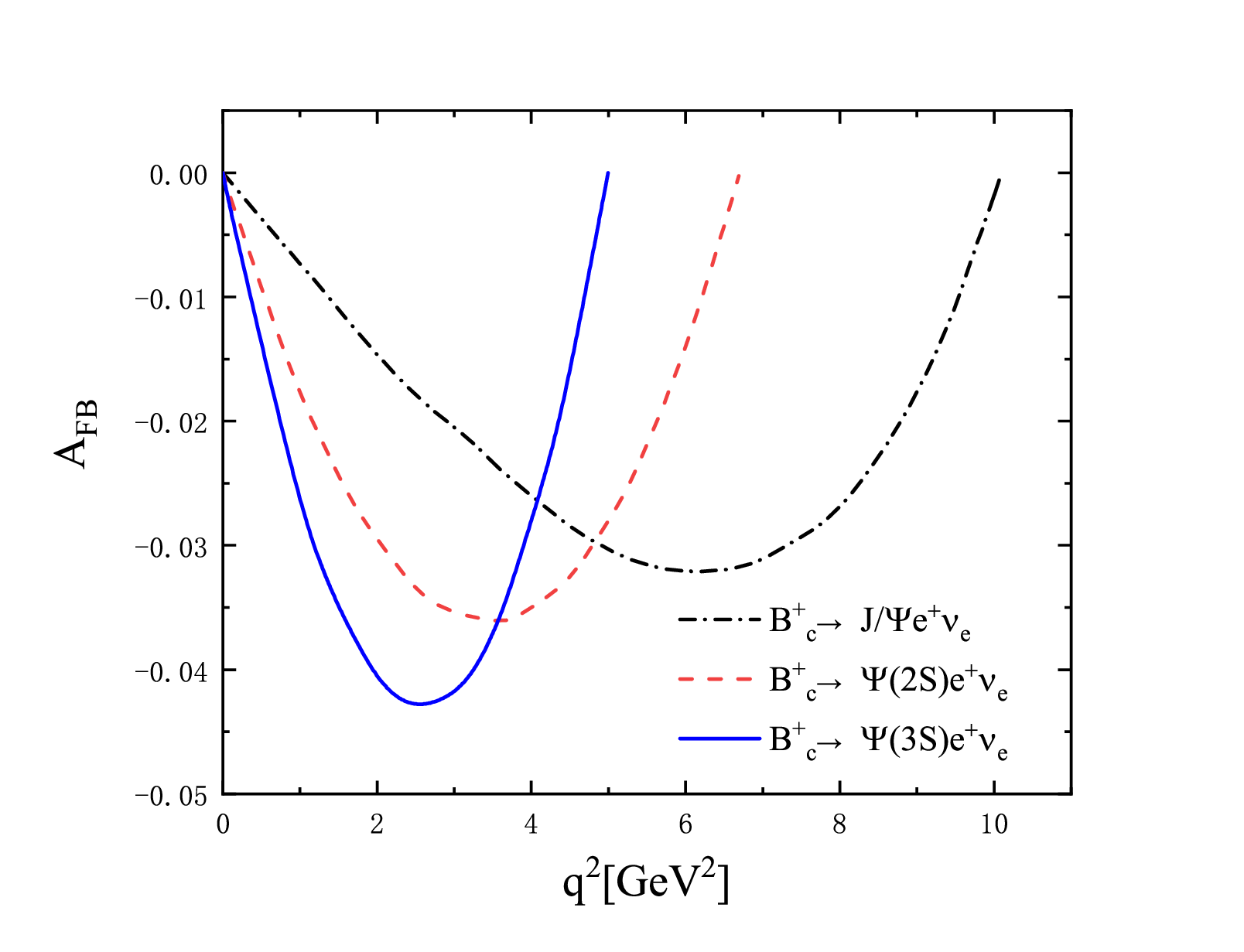}\quad}
  \subfigure[]{\includegraphics[width=0.30\textwidth]{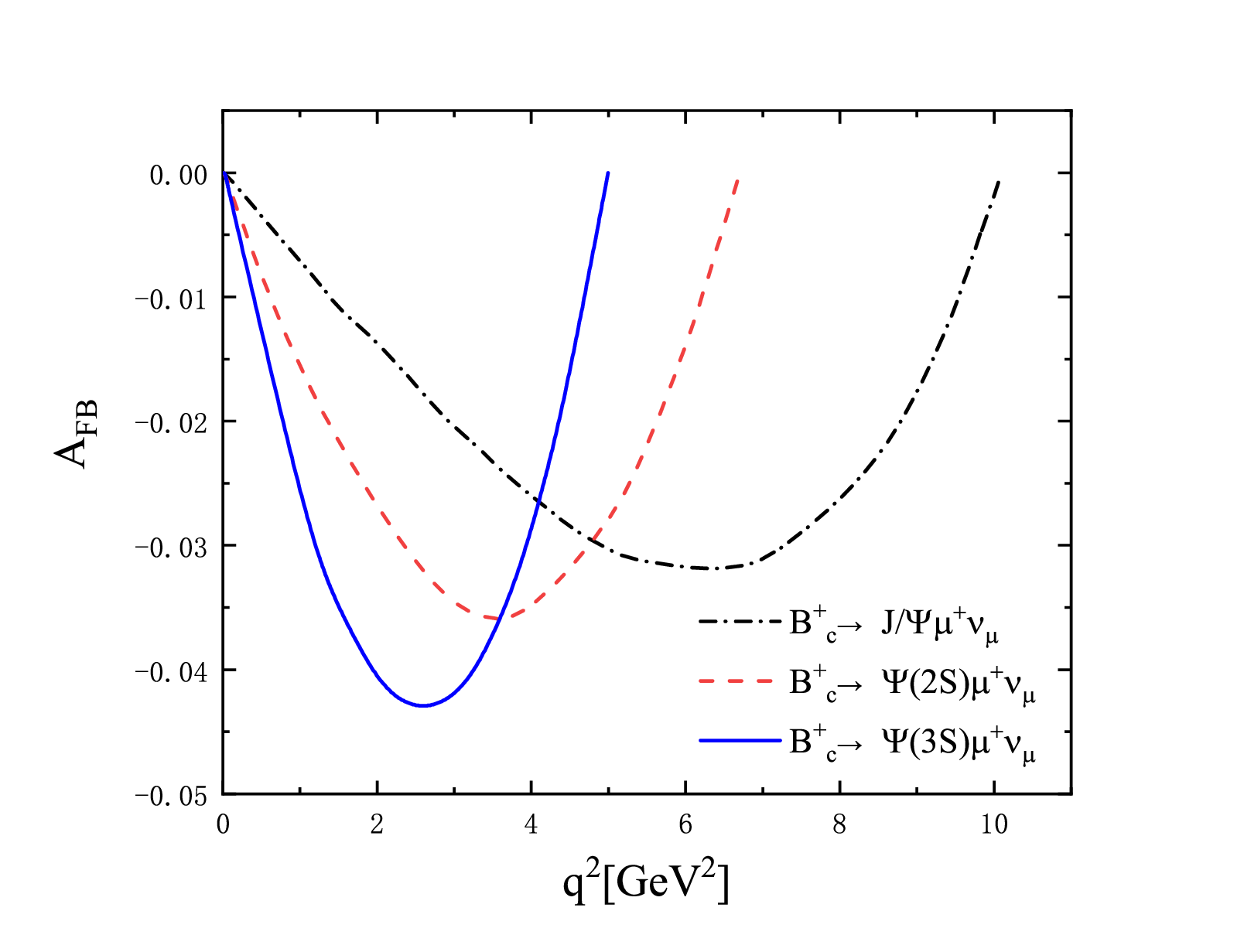}\quad}
  \subfigure[]{\includegraphics[width=0.30\textwidth]{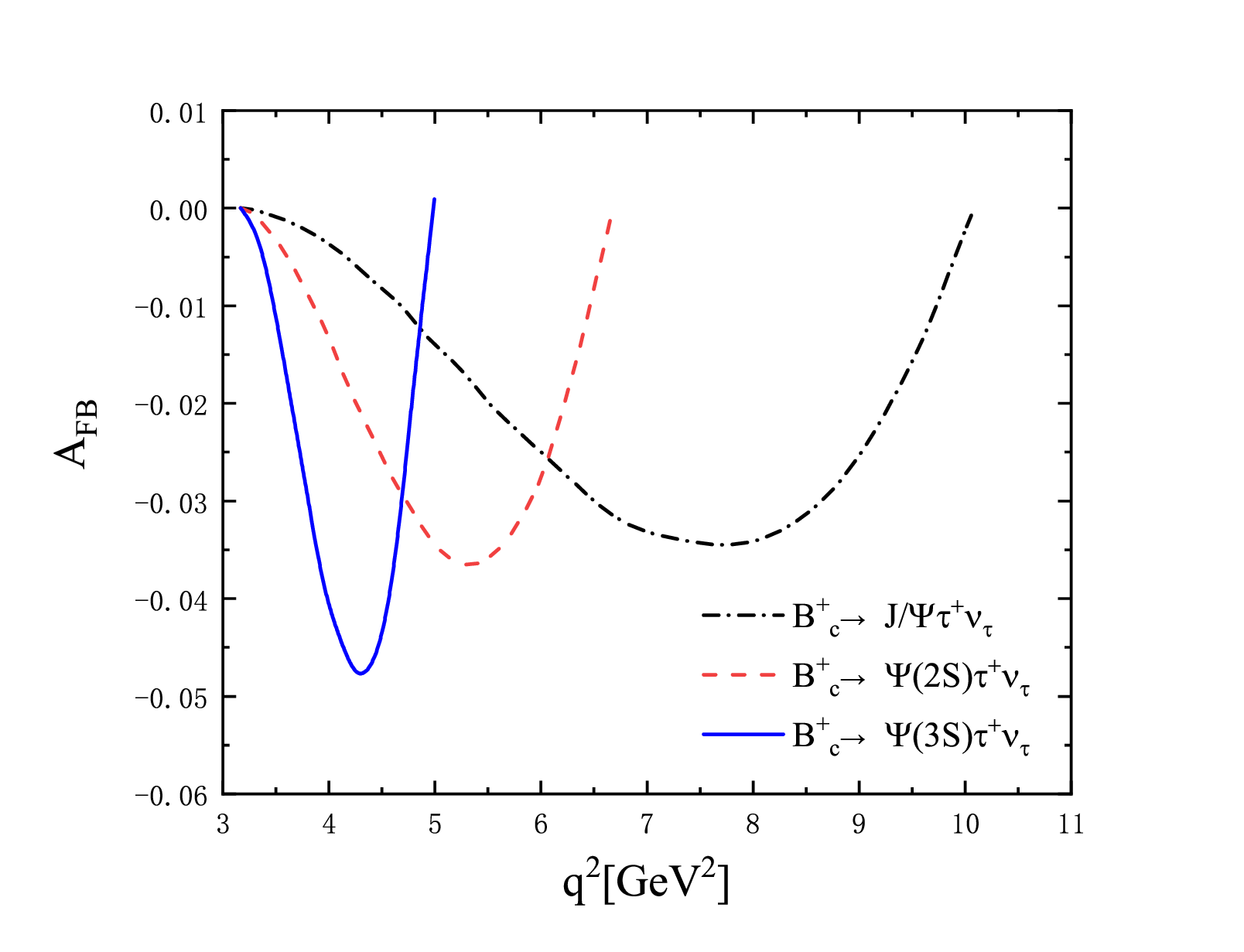}}
\caption{The $q^{2}$ dependences of the forward-backward asymmetries $A_{FB}$ for the decays $B_c\to \eta_c(1S,2S,3S)\ell^+\nu_{\ell}$
and $B_c\to \psi(1S,2S,3S)\ell^+\nu_{\ell}$.}\label{fig:T1}
\end{figure}

In Figs. \ref{fig:T22} and \ref{fig:T1}, we display the  $q^{2}$-dependences of the differential decay rates $d\Gamma_{(L)}/dq^{2}$ and the forward-backward asymmetries $A_{FB}$, respectively. It can be observed that the values of $d\Gamma_{(L)}/dq^{2}$ and $A_{FB}$ coincide with 0 at the zero recoil point $(q^{2}=q^{2}_{max})$ since the coefficient $\lambda(q^2)=\lambda(m^{2}_{B_{c}},m^{2}_{\eta_c/\psi},q^{2})$ shown in Eqs. (\ref{eq:pp})-(\ref{eq:btheta2}) at the same zero recoil point being equal to 0. Furthermore, it is very different for the $q^2$ dependences of the differential decay rates
from the longitudinal polarization $d\Gamma_{L}/dq^{2}$ between the decays $B_c^+\to J/\Psi \ell^{\prime+}\nu_{\ell^\prime}$ and
$B_c^+\to \psi(2S,3S) \ell^{\prime+}\nu_{\ell^\prime}$.


Similarly, we also analysis the semileptonic decays $B_{c}^{+}\to X(3872)\ell^{+}\nu_{\ell}$ by assuming the $X(3872)$ as a regular $c\bar c$ charmonium state. The branching ratios, the longitudinal polarization fractions $f_L$ and the forward-backward asymmetries $A_{FB}$ are listed in Tab. \ref{tab1x2}.
One can find that $\mathcal{B}r(B_{c}^{+}\to X(3872)\tau^{+}\nu_{\tau})$ is only about one twenty-fifth of $\mathcal{B}r(B_{c}^{+}\to X(3872)\ell^{\prime+}\nu_{\ell^\prime})$  due to the narrower phase space of the final states with the $\tau$ lepton involved.
 Obviously, our predictions
 are consistent with those given in the generalized factorization approach (GFA) \cite{Hsiao:2016pml1}, while are about $7\sim9$ times smaller than those in the QCDSR approach \cite{M08}. The ratio $R_X$ is
\be
\label{rx}
R_{X}=\frac{\mathcal{B} r\left(B_{c}^{+} \to X(3872)\tau^{+} \nu_{\tau}\right)}{\mathcal{B} r\left(B_{c}^{+} \to X(3872) \mu^{+} \nu_{\mu}\right)}=0.040\pm0.006,
\en
which is consistent with the value $R_X=0.048$ given by the GFA and QCDSR.
 There exists significant difference between $B_{c}^{+}\to X(3872) \ell^{\prime+}\nu_{\ell^\prime}$  and $B_{c}^{+}\to J/\psi \ell^{+\prime}\nu_{\ell^\prime}$ with in the polarization fractions: the transverse polarization is dominant in the former, while the
 longitudinal and transverse polarizations are comparable in the latter. For the decays $B_{c}^{+}\to X(3872)\tau^{+}\nu_{\tau}$ and $B_{c}^{+}\to J/\Psi\tau^{+}\nu_{\tau}$, both of them are dominated by the transverse polarization. The forward-backward asymmetries $A_{FB}$  for the decays
 $B_{c}^{+}\to X(3872)\ell^{+}\nu_{\ell}$ are about two times larger in magnitude than those for the channels  $B_{c}^{+}\to J/\Psi \ell^{+}\nu_{\ell}$.
 %Meanwhile, $A_{FB}^{e}$ and $A_{FB}^{\mu}$ are nearly equal to each other, which are larger than the value of $A_{FB}^{\tau}$ in magnitude.
 As to the partial branching ratios and longitudinal polarization fractions in Region 1 and Region 2 defined
 in the previous section are given in Tab. \ref{tablex4}, where the transverse polarizations are dominant in both cases.
 It is evident that the dominant contributions arise from the transverse polarizations whether in Region 1, Region 2, or the entire physical region. All of these values are meaningful to determining the inner structure of X(3872) by comparing with the future experimental measurements.

In Fig. \ref{fig:3872}, we plot the $q^{2}$ dependences of the differential decay rates $d\Gamma_{(L)}/dq^{2}$ with L referring the longitudinal polarization and the forward-backward asymmetries $A_{FB}$ for the decays $B_{c}^{+}\to X(3872)\ell^{+}\nu_{\ell}$, where it is similar in character to the cases of
$B_{c}^{+}\to \psi(nS)\ell^{+}\nu_{\ell}$, that is the values of $d\Gamma_{(L)}/dq^{2}$ and $A_{FB}$ are coincide with 0 at the zero recoil point $(q^{2}=q^{2}_{max})$. Furthermore, the differential decay rate from the transverse polarization $d\Gamma_{T}/dq^{2}(B_{c}^{+}\to X(3872)\ell^{+}\nu_{\ell})$ is much more sensitive to the change of $q^2$ compared with that from the longitudinal polarization
$d\Gamma_{L}/dq^{2}(B_{c}^{+}\to X(3872)\ell^{+}\nu_{\ell})$.
\begin{table}[t]
	\renewcommand\arraystretch{1.8}
	\begin{center}
		\caption{\label{tab:input} \small Branching ratios, the longitudinal polarization fractions $f_{L}$ and the forward-backward asymmetries $A_{FB}$ for the semileptonic decays $B_{c}^{+}\to X(3872)\ell^{+}\nu_{\ell}$ .}
		\begin{tabular}{c|c|c|c|c|c}
\hline\hline
			Decay Modes &\multicolumn{2}{c}{$\mathcal{B} r(10^{-4})$} &$$&{$f_{L}(\%)$ }&$A_{FB}$\\
\hline\hline
			$$ & This work & \cite{Hsiao:2016pml1}&\cite{M08} & $$& $$\\
			\hline
			$B_{c}^{+}\to X(3872) e^{+}\nu_{e}$& $9.27^{+0.16+0.01+0.58}_{-0.16-0.22-0.89} $ & $13.5$&$67$ & $36.7^{+0.6+0.0+2.5}_{-0.6-0.1-1.3}$ & $-0.45^{+0.01+0.00+0.07}_{-0.01-0.00-0.05}$\\
			\hline
$B_{c}^{+}\to X(3872) \mu^{+}\nu_{\mu}$&$9.20^{+0.16+0.01+0.57}_{-0.16-0.21-0.89}$&$13.5$&$-$&$36.7^{+0.6+0.0+2.5}_{-0.6-0.0-1.3}$&$-0.44^{+0.01+0.00+0.07}_{-0.01-0.00-0.05}$\\
\hline
$B_{c}^{+}\to X(3872) \tau^{+}\nu_{\tau}$&$0.37^{+0.01+0.00+0.04}_{-0.01-0.00-0.04}$&$0.65$&$3.2$&$31.9^{+0.6+0.1+2.6}_{-0.6-0.5-3.1}$&$-0.28^{+0.01+0.00+0.05}_{-0.01-0.00-0.04}$\\
\hline\hline
\end{tabular}\label{tab1x2}
	\end{center}
\end{table}
\begin{table}[H]
\caption{The partial branching ratios and longitudinal polarization fraction $f_{L}$ (in\%) for the decays $B_{c}^{+}\to X(3872) \ell^{+}\nu_{\ell}$.}
\begin{center}
\scalebox{0.6}{
\begin{tabular}{|c|c|c|c|c|c|c|c|c|}
\hline\hline
Observables&Region 1&Region 2&Observables&Region 1&Region 2&Observables&Region 1&Region 2\\
\hline
$\mathcal{B} r(B_{c}^{+}\to X(3872) e^{+}\nu_{e})$&$0.47\times10^{-3}$&$0.46\times10^{-3}$&$\mathcal{B} r(B_{c}^{+}\to X(3872) \mu^{+}\nu_{\mu})$&$0.46\times10^{-3}$&$0.46\times10^{-3}$&$\mathcal{B} r(B_{c}^{+}\to X(3872) \tau^{+}\nu_{\tau})$&$1.15\times10^{-5}$&$2.58\times10^{-5}$\\
$f_{L}(B_{c}^{+}\to X(3872) e^{+}\nu_{e})$&$43.4$&$29.9$&$f_{L}(B_{c}^{+}\to X(3872) \mu^{+}\nu_{\mu})$&$43.4$&$29.9$&$f_{L}(B_{c}^{+}\to X(3872) \tau^{+}\nu_{\tau})$&$32.3$&$31.7$\\
\hline\hline
\end{tabular}\label{tablex4}}
\end{center}
\end{table}

\begin{figure}[H]
\renewcommand{\arraystretch}{0.5}
\begin{tabular}{cc}
\includegraphics[width=0.5\textwidth]{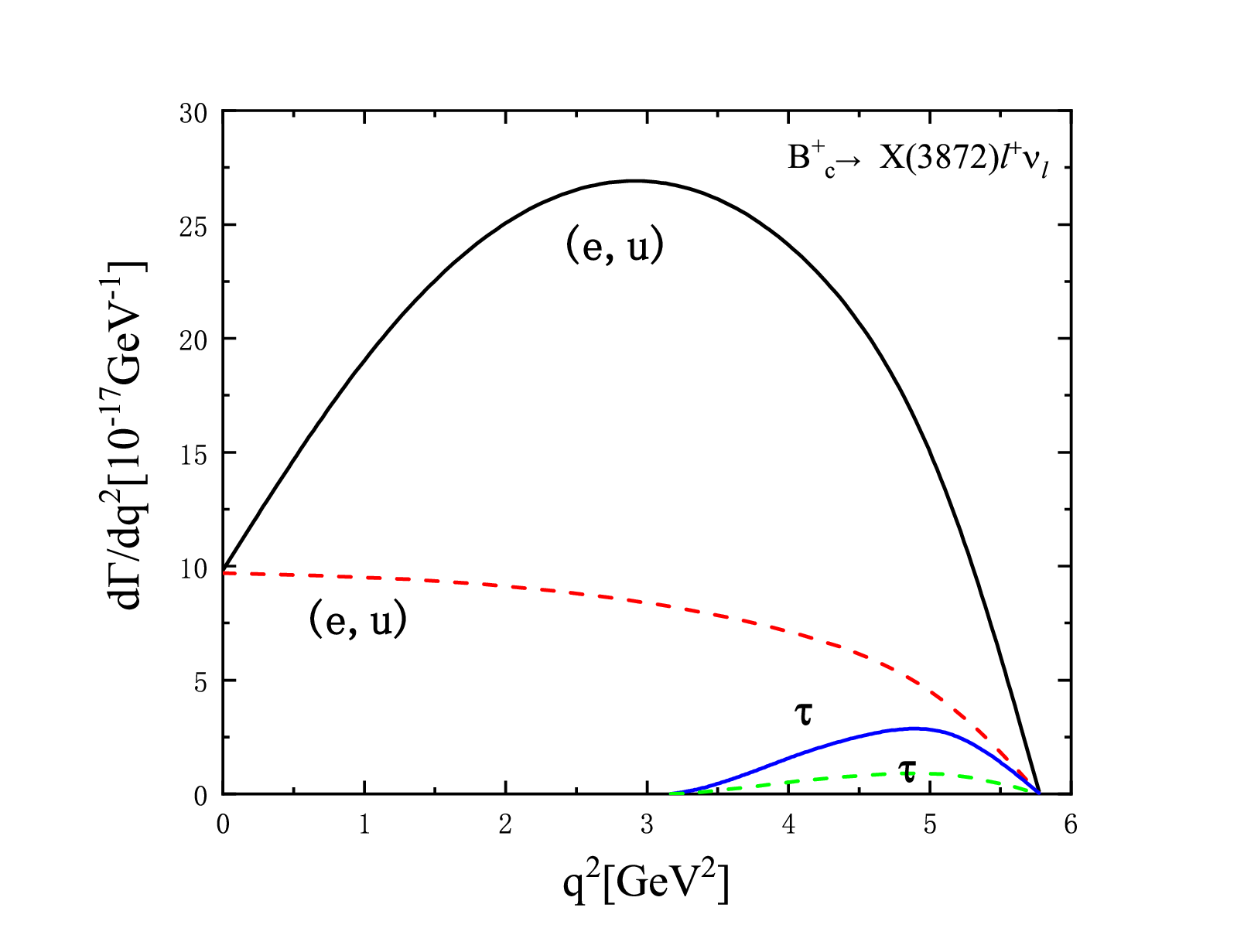}&
\hfill\includegraphics[width=0.5\textwidth]{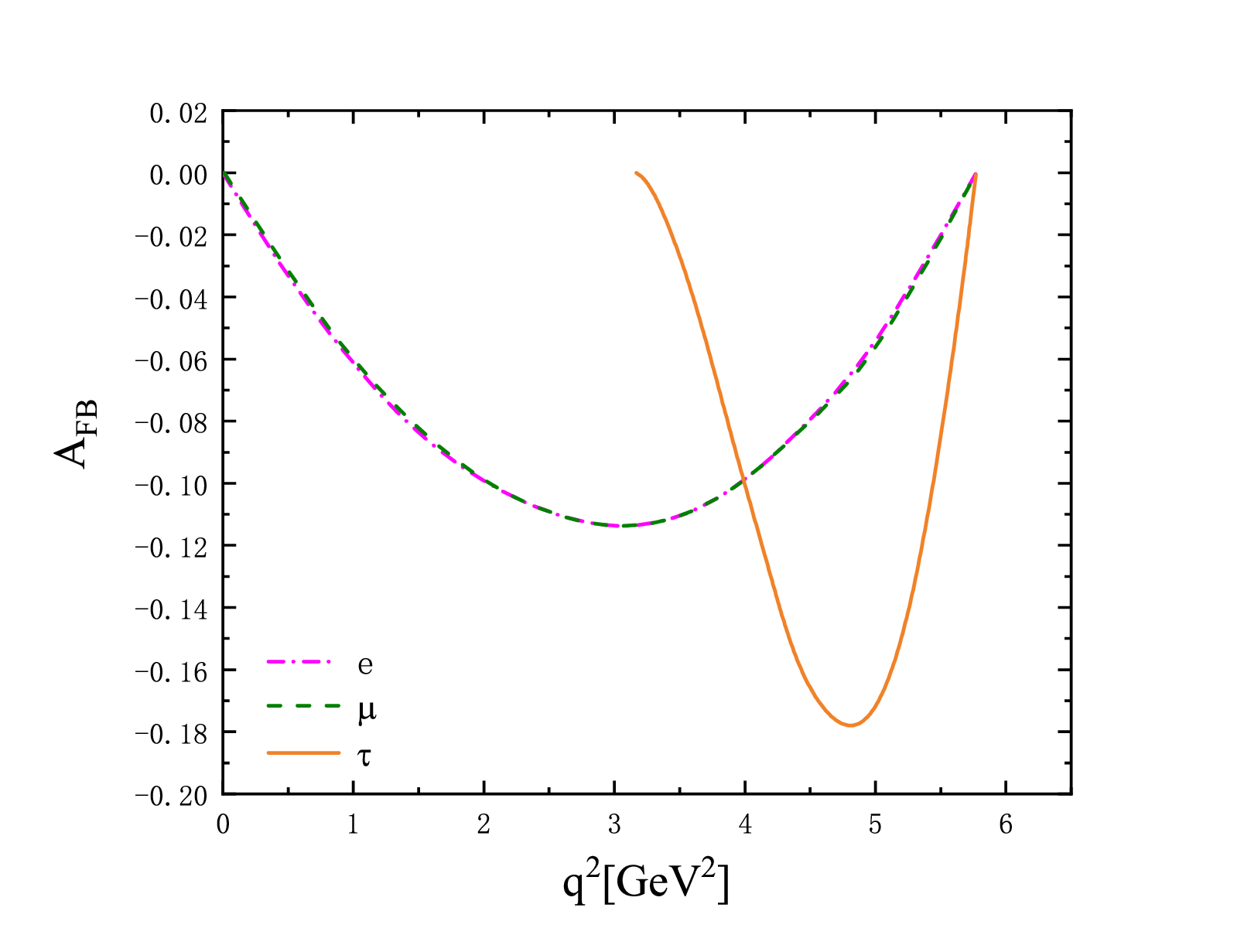}
\end{tabular}
\vspace*{-3mm}
\caption{The $q^{2}$ dependences of the differential decay rates $d\Gamma/dq^{2}$ (solid lines) and $d\Gamma_{L}/dq^{2}$  (dashed lines) are shown on the left pannel and the $q^{2}$ dependences of the forward-backward asymmetries $A_{FB}$ for the decays $B_{c}^{+}\to X(3872) \ell^{+}\nu_{\ell}$ are shown on the right pannel.}
\label{fig:3872}
\end{figure}

\section{Summary}\label{sum}

 In this work, we investigated the exclusive semileptonic $B_{c}$ decays to $\eta_c(1S,2S,3S), \psi(1S,2S,3S)$ and $X(3872)$ in the CLFQM. Using the helicity amplitudes combined via the form factors, we calculated the branching ratios, the longitudinal polarization fractions $f_{L}$ and the forward-backward asymmetries $A_{FB}$ for these semileptonic $B_c$ decays.
From the numerical results, we found the following points:
\begin{enumerate}
\item
Compared the branching ratios of these semileptonic decays, it shows a clear hierarchical relationship,
\be
\mathcal{B} r(B_{c} \to X(3S)\ell\nu_{\ell})< \mathcal{B} r(B_{c} \rightarrow X(2S)\ell\nu_{\ell})< \mathcal{B} r(B_{c} \rightarrow X(1S)\ell\nu_{\ell}),
\en
where $X=\eta_{c}, \psi$.
%\item
%There exists an approximate relation between the branching ratios
%\be
%\mathcal{B}r(B_c\to \psi(nS) \ell\bar\nu_{\ell})\approx 2\mathcal{B}r(B_c\to \eta_c(nS)\ell\bar\nu_{\ell}),
%\en
%with $\ell=e, \mu, \tau$ and $n=1,2,3$.
\item
The ratios $R_X=\frac{\mathcal{B} r\left(B_{c}^{+} \to X\tau^{+} \nu_{\tau}\right)}{\mathcal{B} r\left(B_{c}^{+} \to X\mu^{+} \nu_{\mu}\right)}$ are predicted as
\be
R_{J/\Psi}&=&0.25\pm0.04, R_{\psi(2S)}=0.068\pm0.003, R_{\psi(3S)}=0.0081\pm0.0007,\\
R_{\eta_c}&=&0.29\pm0.02, R_{\eta_c(2S)}=0.09\pm0.12, R_{\eta_c(3S)}=0.012\pm0.006,\\
R_{X(3872)}&=&0.040\pm0.006.
\en
\item
The longitudinal polarization fractions $f_{L}$ of the decays $B_{c}^{+}\to \psi \ell^{+}\nu_{\ell}$ have the following rules
\be
 f_L(B_{c}^{+}\to \psi(nS) e^{+}\nu_{e})\approx f_L(B_{c}^{+}\to \psi(nS) \mu^{+}\nu_{\mu})>f_L(B_{c}^{+}\to \psi(nS) \tau^{+}\nu_{\tau}).
\en
The longitudinal polarization dominates the branching ratios for the decays with $e$ and $\mu$ involved, while the transverse polarization
is dominant for the decays with $\tau$ involved. It is different for the decays $B_{c}^{+}\to X(3872) \ell^{+}\nu_{\ell}$,
where the transverse polarization is dominant in any case.
\item
The ratios of the forward-backward asymmetries $A_{FB}$ between the different semileptonic decays associated with the $B_{c}\rightarrow \eta_{c}(nS)$ transitions are given as
\be
 \frac{A_{FB}^{\mu}}{A_{FB}^{e}}=(3\sim4)\times10^{4}, \;\;\;\;\;\; \frac{A_{FB}^{\tau}}{A_{FB}^{e}} > 1\times10^{5}.
\en
While the forward-backward asymmetries $A_{FB}$ for the decays $B_{c}^{+}\to \psi(nS) \ell^{+}\nu_{\ell}$ and $B_{c}^{+}\to X(3872) \ell^{+}\nu_{\ell}$ become minus in sign, the differences between these values $A_{FB}^{\ell}$ are small and less than 3 times.
\item
Our predictions are helpful to test the standard model and different new physics scenarios. Certainly, the values correlated with the $X(3872)$ are also meaningful to determining the inner structure of the $X(3872)$ by comparing with the future experimental measurements.
\end{enumerate}
%%%%%%%%%%%%%%%%%%%%%%%%%%%%%%%%%%%%%%%%%%%%%%%%%%%%%%%%%%%%%%%%%%%%%%%%%%%%%%%
\section*{Acknowledgment}
This work is partly supported by the National Natural Science
Foundation of China under Grant No. 11347030, the Program of
Science and Technology Innovation Talents in Universities of Henan
Province 14HASTIT037, as well as the Natural Science Foundation of Henan
Province under Grant No. 232300420116.
\appendix
\section{Some specific rules under the $p^-$ intergration}
When preforming the integraion, we need to include the zero-mode contributions. It amounts to performing the integration in a proper way in the CLFQM. Specificlly we
use the following rules given in Refs. \cite{jaus,hycheng}
\be \hat{p}_{1 \mu}^{\prime} &\doteq &   P_{\mu}
A_{1}^{(1)}+q_{\mu} A_{2}^{(1)},\\
\hat{p}_{1 \mu}^{\prime}
\hat{p}_{1 \nu}^{\prime}  &\doteq & g_{\mu \nu} A_{1}^{(2)} +P_{\mu}
P_{\nu} A_{2}^{(2)}+\left(P_{\mu} q_{\nu}+q_{\mu} P_{\nu}\right)
A_{3}^{(2)}+q_{\mu} q_{\nu} A_{4}^{(2)},\\
Z_{2}&=&\hat{N}_{1}^{\prime}+m_{1}^{\prime 2}-m_{2}^{2}+\left(1-2
x_{1}\right) M^{\prime 2} +\left(q^{2}+q \cdot P\right)
\frac{p_{\perp}^{\prime} \cdot q_{\perp}}{q^{2}},\\
A_{1}^{(1)}&=&\frac{x_{1}}{2}, \quad A_{2}^{(1)}=
A_{1}^{(1)}-\frac{p_{\perp}^{\prime} \cdot q_{\perp}}{q^{2}},\quad A_{3}^{(2)}=A_{1}^{(1)} A_{2}^{(1)},\\
A_{4}^{(2)}&=&\left(A_{2}^{(1)}\right)^{2}-\frac{1}{q^{2}}A_{1}^{(2)},\quad A_{1}^{(2)}=-p_{\perp}^{\prime 2}-\frac{\left(p_{\perp}^{\prime}
\cdot q_{\perp}\right)^{2}}{q^{2}}, \quad A_{2}^{(2)}=\left(A_{1}^{(1)}\right)^{2}.  \en
\section{EXPRESSIONS OF $B_{c} \rightarrow P,V,A$ FORM FACTORS}
The following are the analytical expressions of the $B_{c} \to \eta_{c}(1S, 2S, 3S)$, $\psi(1S,2S,3S),X(3872)$ transition form factors in the covariant light-front quark model
\begin{footnotesize}
\begin{eqnarray}
F^{B_c\eta_c}_{1}\left(q^{2}\right)&=&\frac{N_{c}}{16 \pi^{3}} \int d x_{2} d^{2} p_{\perp}^{\prime} \frac{h_{B_c}^{\prime}
h_{\eta_c}^{\prime \prime}}{x_{2} \hat{N}_{1}^{\prime} \hat{N}_{1}^{\prime \prime}}\left[x_{1}\left(M_{0}^{\prime 2}+M_{0}^{\prime \prime 2}\right)+x_{2} q^{2}\right.-x_{2}\left(m_{1}^{\prime}-m_{1}^{\prime \prime}\right)^{2}\non
&&-x_{1}\left(m_{1}^{\prime}-m_{2}\right)^{2}-x_{1}\left(m_{1}^{\prime \prime}-m_{2}\right)^{2}]\\
F^{B_c\eta_c}_{0}\left(q^{2}\right)&=&F^{B_c\eta_c}_{1}(q^{2})+\frac{q^2}{(q\cdot P)}\frac{N_{c}}{16 \pi^{3}}  \int d x_{2} d^{2} p_{\perp}^{\prime} \frac{2 h_{B_c}^{\prime}
h_{\eta_c}^{\prime \prime}}{x_{2} \hat{N}_{1}^{\prime} \hat{N}_{1}^{\prime \prime}}\left\{-x_{1} x_{2} M^{\prime 2}
-p_{\perp}^{\prime 2}-m_{1}^{\prime} m_{2}\right.\notag\\
&&+\left(m_{1}^{\prime \prime}-m_{2}\right)\left(x_{2} m_{1}^{\prime}
+x_{1} m_{2}\right)+2 \frac{q \cdot P}{q^{2}}\left(p_{\perp}^{\prime 2}+2 \frac{\left(p_{\perp}^{\prime} \cdot q_{\perp}\right)^{2}}{q^{2}}\right)
+2 \frac{\left(p_{\perp}^{\prime} \cdot q_{\perp}\right)^{2}}{q^{2}}\notag\\
&&\left.-\frac{p_{\perp}^{\prime} \cdot q_{\perp}}{q^{2}}\left[M^{\prime \prime 2}-x_{2}\left(q^{2}+q \cdot P\right)-\left(x_{2}-x_{1}\right) M^{\prime 2}+2 x_{1} M_{0}^{\prime 2}\right.\right.\notag\\
&&\left.\left.-2\left(m_{1}^{\prime}-m_{2}\right)\left(m_{1}^{\prime}+m_{1}^{\prime \prime}\right)\right]\right\},\\
V^{B_c \psi}(q^{2})&=&\frac{N_{c}(M^{'}+M^{''})}{16 \pi^{3}} \int d x_{2} d^{2} p_{\perp}^{\prime} \frac{2 h_{B_c}^{\prime}
 h_{\psi}^{\prime \prime}}{x_{2} \hat{N}_{1}^{\prime} \hat{N}_{1}^{\prime \prime}}\left\{x_{2} m_{1}^{\prime}
 +x_{1} m_{2}+\left(m_{1}^{\prime}-m_{1}^{\prime \prime}\right) \frac{p_{\perp}^{\prime} \cdot q_{\perp}}{q^{2}}\right.\non &&\left.
 +\frac{2}{w_{V}^{\prime \prime}}\left[p_{\perp}^{\prime 2}+\frac{\left(p_{\perp}^{\prime} \cdot q_{\perp}\right)^{2}}{q^{2}}\right]\right\},\\
A_1^{B_c \psi}(q^{2})&=& -\frac{1}{M^{'}+M^{''}}\frac{N_{c}}{16 \pi^{3}} \int d x_{2} d^{2} p_{\perp}^{\prime} \frac{h_{B_c}^{\prime} h_{\psi}^{\prime \prime}}{x_{2}
\hat{N}_{1}^{\prime}
\hat{N}_{1}^{\prime \prime}}\left\{2 x_{1}\left(m_{2}-m_{1}^{\prime}\right)\left(M_{0}^{\prime 2}+M_{0}^{\prime \prime 2}\right)
-4 x_{1} m_{1}^{\prime \prime} M_{0}^{\prime 2}\right.\non
&&\left.+2 x_{2} m_{1}^{\prime} q \cdot P+2 m_{2} q^{2}-2 x_{1} m_{2}\left(M^{\prime 2}+M^{\prime \prime 2}\right)+2\left(m_{1}^{\prime}-m_{2}\right)\left(m_{1}^{\prime}
+m_{1}^{\prime \prime}\right)^{2}+8\left(m_{1}^{\prime}-m_{2}\right) \right.\non &&
\left. \times\left[p_{\perp}^{\prime 2}+\frac{\left(p_{\perp}^{\prime}
\cdot q_{\perp}\right)^{2}}{q^{2}}\right]+2\left(m_{1}^{\prime}+m_{1}^{\prime \prime}\right)\left(q^{2}+q \cdot P\right) \frac{p_{\perp}^{\prime} \cdot q_{\perp}}{q^{2}}
-4 \frac{q^{2} p_{\perp}^{\prime 2}+\left(p_{\perp}^{\prime} \cdot q_{\perp}\right)^{2}}{q^{2} w_{\psi}^{\prime \prime}}
\right.\non && \left.\times\left[2 x_{1}\left(M^{\prime 2}+M_{0}^{\prime 2}\right)-q^{2}-q \cdot P-2\left(q^{2}+q \cdot P\right) \frac{p_{\perp}^{\prime} \cdot q_{\perp}}{q^{2}}-2\left(m_{1}^{\prime}-m_{1}^{\prime \prime}\right)\left(m_{1}^{\prime}-m_{2}\right)\right]\right\},\;\;\;\;\;\\
A_2^{B_c \psi}(q^{2})&=& \frac{N_{c}(M^{'}+M^{''})}{16 \pi^{3}} \int d x_{2} d^{2} p_{\perp}^{\prime} \frac{2 h_{B_c}^{\prime} h_{\psi}^{\prime \prime}}{x_{2} \hat{N}_{1}^{\prime}
\hat{N}_{1}^{\prime \prime}}\left\{\left(x_{1}-x_{2}\right)\left(x_{2} m_{1}^{\prime}+x_{1} m_{2}\right)-\frac{p_{\perp}^{\prime} \cdot q_{\perp}}{q^{2}}\left[2 x_{1} m_{2}
+m_{1}^{\prime \prime} \right.\right.\non &&
\left.\left.+\left(x_{2}-x_{1}\right) m_{1}^{\prime}\right]-2 \frac{x_{2} q^{2}+p_{\perp}^{\prime} \cdot q_{\perp}}{x_{2} q^{2} w_{\psi}^{\prime \prime}}\left[p_{\perp}^{\prime} \cdot p_{\perp}^{\prime \prime}
+\left(x_{1} m_{2}+x_{2} m_{1}^{\prime}\right)\left(x_{1} m_{2}-x_{2} m_{1}^{\prime \prime}\right)\right]\right\},\\
A_0^{B_c \psi}(q^{2})&=& \frac{M^{'}+M^{''}}{2M^{''}}A_1^{B_c \psi}(q^{2})-\frac{M^{'}-M^{''}}{2M^{''}}A_2^{B_c \psi}(q^{2})-\frac{q^2}{2M^{''}}\frac{N_{c}}{16 \pi^{3}} \int d x_{2} d^{2} p_{\perp}^{\prime} \frac{h_{B_c}^{\prime} h_{\psi}^{\prime \prime}}{x_{2} \hat{N}_{1}^{\prime}
\hat{N}_{1}^{\prime \prime}}\left\{2\left(2 x_{1}-3\right)\right.\non &&\left.\left(x_{2} m_{1}^{\prime}+x_{1} m_{2}\right)-8\left(m_{1}^{\prime}-m_{2}\right)
\times\left[\frac{p_{\perp}^{\prime 2}}{q^{2}}
+2 \frac{\left(p_{\perp}^{\prime} \cdot q_{\perp}\right)^{2}}{q^{4}}\right]-\left[\left(14-12 x_{1}\right) m_{1}^{\prime}\right.\right. \non &&\left.\left.-2 m_{1}^{\prime \prime}-\left(8-12 x_{1}\right) m_{2}\right] \frac{p_{\perp}^{\prime} \cdot q_{\perp}}{q^{2}}
+\frac{4}{w_{\psi}^{\prime \prime}}\left(\left[M^{\prime 2}+M^{\prime \prime 2}-q^{2}+2\left(m_{1}^{\prime}-m_{2}\right)\left(m_{1}^{\prime \prime}
+m_{2}\right)\right]\right.\right.\non &&\left.\left.\times\left(A_{3}^{(2)}+A_{4}^{(2)}-A_{2}^{(1)}\right)
+Z_{2}\left(3 A_{2}^{(1)}-2 A_{4}^{(2)}-1\right)+\frac{1}{2}\left[x_{1}\left(q^{2}+q \cdot P\right)
-2 M^{\prime 2}-2 p_{\perp}^{\prime} \cdot q_{\perp}\right.\right.\right.\non &&\left.\left.\left.-2 m_{1}^{\prime}\left(m_{1}^{\prime \prime}+m_{2}\right)
-2 m_{2}\left(m_{1}^{\prime}-m_{2}\right)\right]\left(A_{1}^{(1)}+A_{2}^{(1)}-1\right) q \cdot P\left[\frac{p_{\perp}^{\prime 2}}{q^{2}}
+\frac{\left(p_{\perp}^{\prime} \cdot q_{\perp}\right)^{2}}{q^{4}}\right]\right.\right.\non &&\left.\left.\times\left(4 A_{2}^{(1)}-3\right)\right)\right\},\;\;\;
\end{eqnarray}
\end{footnotesize}
\begin{footnotesize}
\begin{eqnarray}
A^{B_c X}(q^{2})&=&(M^{'}-M^{''})\frac{N_{c}}{16 \pi^{3}} \int d x_{2} d^{2} p_{\perp}^{\prime} \frac{2 h_{B_c}^{\prime}
 h_{X}^{\prime \prime}}{x_{2} \hat{N}_{1}^{\prime} \hat{N}_{1}^{\prime \prime}}\left\{x_{2} m_{1}^{\prime}
 +x_{1} m_{2}+\left(m_{1}^{\prime}+m_{1}^{\prime \prime}\right) \frac{p_{\perp}^{\prime} \cdot q_{\perp}}{q^{2}}\right.\non &&\left.
 +\frac{2}{w_{X}^{\prime \prime}}\left[p_{\perp}^{\prime 2}+\frac{\left(p_{\perp}^{\prime} \cdot q_{\perp}\right)^{2}}{q^{2}}\right]\right\},\\
V_1^{B_c X}(q^{2})&=& -\frac{1}{M^{'}-M^{''}}\frac{N_{c}}{16 \pi^{3}} \int d x_{2} d^{2} p_{\perp}^{\prime} \frac{h_{B_c}^{\prime} h_{X}^{\prime \prime}}{x_{2}
\hat{N}_{1}^{\prime}
\hat{N}_{1}^{\prime \prime}}\{2 x_{1}\left(m_{2}-m_{1}^{\prime}\right)\left(M_{0}^{\prime 2}+M_{0}^{\prime \prime 2}\right)
+4 x_{1} m_{1}^{\prime \prime} M_{0}^{\prime 2}\non&&+2 x_{2} m_{1}^{\prime} q \cdot P
\left.+2 m_{2} q^{2}-2 x_{1} m_{2}\left(M^{\prime 2}+M^{\prime \prime 2}\right)+2\left(m_{1}^{\prime}-m_{2}\right)\left(m_{1}^{\prime}
-m_{1}^{\prime \prime}\right)^{2}+8\left(m_{1}^{\prime}-m_{2}\right) \right.\non &&
\left. \times\left[p_{\perp}^{\prime 2}+\frac{\left(p_{\perp}^{\prime}
\cdot q_{\perp}\right)^{2}}{q^{2}}\right]+2\left(m_{1}^{\prime}-m_{1}^{\prime \prime}\right)\left(q^{2}+q \cdot P\right) \frac{p_{\perp}^{\prime} \cdot q_{\perp}}{q^{2}}
-4 \frac{q^{2} p_{\perp}^{\prime 2}+\left(p_{\perp}^{\prime} \cdot q_{\perp}\right)^{2}}{q^{2} w_{X}^{\prime \prime}}
\right.\non && \left.\times\left[2 x_{1}\left(M^{\prime 2}+M_{0}^{\prime 2}\right)-q^{2}-q \cdot P-2\left(q^{2}+q \cdot P\right) \frac{p_{\perp}^{\prime} \cdot q_{\perp}}{q^{2}}-2\left(m_{1}^{\prime}+m_{1}^{\prime \prime}\right)\left(m_{1}^{\prime}-m_{2}\right)\right]\right\},\;\;\;\\
V_2^{B_c X}(q^{2})&=& (M^{'}-M^{''})\frac{N_{c}}{16 \pi^{3}} \int d x_{2} d^{2} p_{\perp}^{\prime} \frac{2 h_{B_c}^{\prime} h_{X}^{\prime \prime}}{x_{2} \hat{N}_{1}^{\prime}
\hat{N}_{1}^{\prime \prime}}\left\{(x_{1}-x_{2}\right)\left(x_{2} m_{1}^{\prime}+x_{1} m_{2}\right)-[2 x_{1} m_{2}
-m_{1}^{\prime \prime}\non &&+\left(x_{2}-x_{1}\right) m_{1}^{\prime}]
\times \frac{p_{\perp}^{\prime} \cdot q_{\perp}}{q^{2}}-2 \frac{x_{2} q^{2}+p_{\perp}^{\prime} \cdot q_{\perp}}{x_{2} q^{2} w_{X}^{\prime \prime}}[p_{\perp}^{\prime} \cdot p_{\perp}^{\prime \prime}
+\left(x_{1} m_{2}+x_{2} m_{1}^{\prime}\right)\non &&\times\left(x_{1} m_{2}+x_{2} m_{1}^{\prime \prime}\right)]\},\\
V_0^{B_c X}(q^{2})&=& \frac{M^{'}-M^{''}}{2M^{''}}V_1^{B_c X}(q^{2})-\frac{M^{'}+M^{''}}{2M^{''}}V_2^{B_c }(q^{2})-\frac{q^2}{2M^{''}}\frac{N_{c}}{16 \pi^{3}} \int d x_{2} d^{2} p_{\perp}^{\prime} \frac{h_{B_c}^{\prime} h_{X}^{\prime \prime}}{x_{2} \hat{N}_{1}^{\prime}
\hat{N}_{1}^{\prime \prime}}\non &&\times\{2\left(2 x_{1}-3\right)\left(x_{2} m_{1}^{\prime}+x_{1} m_{2}\right)-8\left(m_{1}^{\prime}-m_{2}\right)
\left[\frac{p_{\perp}^{\prime 2}}{q^{2}}
+2 \frac{\left(p_{\perp}^{\prime} \cdot q_{\perp}\right)^{2}}{q^{4}}\right]-[\left(14-12 x_{1}\right) m_{1}^{\prime}\non &&+2 m_{1}^{\prime \prime}-\left(8-12 x_{1}\right) m_{2}] \frac{p_{\perp}^{\prime} \cdot q_{\perp}}{q^{2}}
+\frac{4}{w_{X}^{\prime \prime}}(\left[M^{\prime 2}+M^{\prime \prime 2}-q^{2}+2\left(m_{1}^{\prime}-m_{2}\right)\left(-m_{1}^{\prime \prime}
+m_{2}\right)\right]\non &&\times\left(A_{3}^{(2)}+A_{4}^{(2)}-A_{2}^{(1)}\right)
+Z_{2}\left(3 A_{2}^{(1)}-2 A_{4}^{(2)}-1\right)+\frac{1}{2}[x_{1}\left(q^{2}+q \cdot P\right)
-2 M^{\prime 2}-2 p_{\perp}^{\prime} \cdot q_{\perp}\non &&-2 m_{1}^{\prime}\left(-m_{1}^{\prime \prime}+m_{2}\right)
\left.-2 m_{2}\left(m_{1}^{\prime}-m_{2}\right)\right]\left(A_{1}^{(1)}+A_{2}^{(1)}-1\right)\non &&
\left.\left.\times q \cdot P\left[\frac{p_{\perp}^{\prime 2}}{q^{2}}
+\frac{\left(p_{\perp}^{\prime} \cdot q_{\perp}\right)^{2}}{q^{4}}\right]\left(4 A_{2}^{(1)}-3\right)\right)\right\}.\;\;\;
\end{eqnarray}
\end{footnotesize}

%%%%%%%%%%%%%%%%%%%%%%%%%%%%%%%%%%%%%%%%%%%%%%%%%%%%%%%%%%%%%%%%%%%%%%%%
%                               references
%%%%%%%%%%%%%%%%%%%%%%%%%%%%%%%%%%%%%%%%%%%%%%%%%%%%%%%%%%%%%%%%%%%%%%%%


\begin{thebibliography}{99}
\bibitem{tere}
 M. V. Terentev, Sov. J. Nucl. Phys. {\bf24}, 106 (1976) [Yad. Fiz. {\bf24}, 207 (1976)].
\bibitem{bere}
 V. B. Berestetsky and M. V. Terentev, Sov. J. Nucl. Phys. {\bf25}, 347 (1977) [Yad. Fiz. {\bf25}, 653 (1977)].
\bibitem{DiracO}
P. A. Dirac, Rev. Mod. Phys. {\bf21}, 392 (1949).

\bibitem{jaus0}
W. Jaus,  \prd {\bf41}, 3394 (1990).

\bibitem{jaus1}
W. Jaus,  \prd {\bf44}, 2851 (1991).

\bibitem{Choi}
H.~M.~Choi and C.~R.~Ji, Phys. Lett. B \textbf{460}, 461 (1999) [arXiv:hep-ph/9903496].

\bibitem{jaus}
W. Jaus, \prd {\bf60}, 054026 (1999).

\bibitem{Zhang:2023ypl}
Z.~Q.~Zhang, Z.~J.~Sun, Y.~C.~Zhao, Y.~Y.~Yang and Z.~Y.~Zhang, Eur. Phys. J. C \textbf{83}, 477 (2023) [arXiv:2301.11107 [hep-ph]].

\bibitem{Wang09}
W.~Wang, Y.~L.~Shen and C.~D.~Lu, Phys. Rev. D {\bf79}, 054012 (2009) [arXiv:0811.3748 [hep-ph]].

\bibitem{Ke14}
H.~W.~Ke, T.~Liu and X.~Q.~Li, Phys. Rev. D \textbf{89}, 017501 (2014) [arXiv:1307.5925 [hep-ph]].

\bibitem{hycheng}
H.~Y.~Cheng, C.~K.~Chua and C.~W.~Hwang, Phys. Rev. D \textbf{69} (2004), 074025 [arXiv:hep-ph/0310359].

\bibitem{Wang:2009mi}
X.~X.~Wang, W.~Wang and C.~D.~Lu, Phys. Rev. D \textbf{79}, 114018 (2009) [arXiv:0901.1934 [hep-ph]].

\bibitem{LHCb:2017vlu}
R.~Aaij \textit{et al.} [LHCb Collaboration], Phys. Rev. Lett. \textbf{120}, 121801 (2018) [arXiv:1711.05623 [hep-ex]].

\bibitem{Cohen:2018dgz}
T.~D.~Cohen, H.~Lamm and R.~F.~Lebed, JHEP \textbf{09}, 168 (2018) [arXiv:1807.02730 [hep-ph]].

\bibitem{AA}
A.~Issadykov and M.~A.~Ivanov, Phys. Lett. B \textbf{783}, 178 (2018) [arXiv:1804.00472 [hep-ph]].

\bibitem{C.F131}
C.~F.~Qiao and R.~L.~Zhu, Phys. Rev. D \textbf{87}, 014009  (2013) [arXiv:1208.5916 [hep-ph]].

\bibitem{Qiao:132}
C.~F.~Qiao, P.~Sun, D.~Yang and R.~L.~Zhu, Phys. Rev. D \textbf{89}, 034008 (2014) [arXiv:1209.5859 [hep-ph]].

\bibitem{R.R}
C.~H. Chang, H.~F.~Fu, G.~L.~Wang and J.~M.~Zhang, Sci. China Phys. Mech. Astron. \textbf{58}, 071001 (2015) [arXiv:1411.3428 [hep-ph]].

\bibitem{ZZT}
T. Zhou, T. h. Wang, Y. Jiang, L. Huo and G. L. Wang, J. Phys. G {\bf48}, 055006 (2021) [arXiv: 2006.05704 [hep-ph]].

\bibitem{Ebert10}
D.~Ebert, R.~N.~Faustov and V.~O.~Galkin, Phys. Rev. D \textbf{68}, 094020 (2003) [arXiv:hep-ph/0306306].

\bibitem{Ebert101}
D.~Ebert, R.~N.~Faustov and V.~O.~Galkin, Phys. Rev. D \textbf{82}, 034019  (2010) [arXiv:1007.1369 [hep-ph]].

\bibitem{Ivanov06}
M.~A.~Ivanov, J.~G.~Korner and P.~Santorelli, Phys. Rev. D \textbf{73}, 054024  (2006) [arXiv:hep-ph/0602050].

\bibitem{M08}
Y.~M.~Wang and C.~D.~Lu, Phys. Rev. D \textbf{77}, 054003  (2008) [arXiv:0707.4439 [hep-ph]].

\bibitem{Huang07}
T. Huang and F. Zuo, Eur. Phys. J. C {\bf51}, 833 (2007) [arXiv:hep-ph/0702147].

\bibitem{Anisimov:1998uk}
A.~Y.~Anisimov, I.~M.~Narodetsky, C.~Semay and B.~Silvestre-Brac, Phys. Lett. B \textbf{452}, 129 (1999) [arXiv:hep-ph/9812514].

\bibitem{Anisimov:1998uk1}
A.~Y.~Anisimov, P.~Y.~Kulikov, I.~M.~Narodetsky and K.~A.~Ter-Martirosian, Phys. Atom. Nucl. \textbf{62}, 1739 (1999) [arXiv:hep-ph/9809249].

\bibitem{Hernndez06}
E.~Hernandez, J.~Nieves and J.~M.~Verde-Velasco, Phys. Rev. D \textbf{74}, 074008  (2006) [arXiv:hep-ph/0607150].

\bibitem{P00}
P.~Colangelo and F.~De Fazio, Phys. Rev. D \textbf{61}, 034012  (2000) [arXiv:hep-ph/9909423].

 \bibitem{I}
I.~Bediaga and J.~H.~Munoz, [arXiv:1102.2190 [hep-ph]].

\bibitem{Rui:2016opu}
Z.~Rui, H.~Li, G.~x.~Wang and Y.~Xiao, Eur. Phys. J. C \textbf{76}, 564 (2016) [arXiv:1602.08918 [hep-ph]].

\bibitem{Wang:2012lrc}
W.~F.~Wang, Y.~Y.~Fan and Z.~J.~Xiao, Chin. Phys. C \textbf{37}, 093102 (2013) [arXiv:1212.5903 [hep-ph]].

\bibitem{JFSD}
J.~F.~Sun, D.~S.~Du and Y.~L.~Yang, Eur. Phys. J. C \textbf{60}, 107 (2009) [arXiv:0808.3619 [hep-ph]].

\bibitem{Ivanov:2000aj}
M.~A.~Ivanov, J.~G.~Korner and P.~Santorelli, Phys. Rev. D \textbf{63}, 074010 (2001) [arXiv:hep-ph/0007169].

\bibitem{Kiselev:2000pp}
V.~V.~Kiselev, A.~E.~Kovalsky and A.~K.~Likhoded, Nucl. Phys. B \textbf{585}, 353 (2000) [arXiv:hep-ph/0002127].

\bibitem{Issadykov:2017wlb}
A.~Issadykov, M.~A.~Ivanov and G.~Nurbakova, EPJ Web Conf. \textbf{158}, 03002 (2017) [arXiv:1907.13210 [hep-ph]].

\bibitem{Nayak:2022gdo}
L.~Nayak, P.~C.~Dash, S.~Kar and N.~Barik, Eur. Phys. J. C \textbf{82}, 750 (2022) [arXiv:2204.04453 [hep-ph]].

\bibitem{zhang}
Z. Q. Zhang, Z. L. Guan, Y. C. Zhao, Z. Y. Zhang, Z. J. Sun, N. Wang and X. D. Ren, Chin. Phys. C {\bf47}, 013103 (2023) [arXiv:2208.07990 [hep-ph]].

\bibitem{ptau3}
Y.~Sakaki, M.~Tanaka, A.~Tayduganov and R.~Watanabe, Phys. Rev. D \textbf{88}, 094012 (2013) [arXiv:1309.0301 [hep-ph]].
\bibitem{pdg}
R. L. Workman \textit{et al.} [Particle Data Group], Review of Particle
Physics, PTEP {\bf2022}, 083C01 (2022).
\bibitem{Berns:2018vpl}
A.~Berns and H.~Lamm, JHEP \textbf{12}, 114 (2018) [arXiv:1808.07360 [hep-ph]].

\bibitem{Lamm:2018xmc}
H.~Lamm [arXiv:1809.08227 [hep-ph]].

\bibitem{Cerri:2018ypt}
A.~Cerri, \textit{et al.},
CERN Yellow Rep. Monogr. \textbf{7}, 867  (2019) [arXiv:1812.07638 [hep-ph]].

\bibitem{ptau2}
M. Tanaka and R. Watanabe,
\prd {\bf 87}, 034028~(2013) [arXiv:1212.1878 [hep-ph]].

\bibitem{prd072012}
J. P. Lees \textit{et al.} [BaBar Collaboration],
\prd {\bf 88}, 072012~(2013) [arXiv:1303.0571 [hep-ex]].

\bibitem{prd114022}
M. A. Ivanov, J. G. Korner and C. T. Tran,
\prd {\bf 92}, 114022~(2015) [arXiv:1508.02678 [hep-ph]].

\bibitem{prd036021}
M. A. Ivanov, J. G. Korner and C. T. Tran, \prd {\bf 95}, 036021~(2017) [arXiv:1701.02937 [hep-ph]].

\bibitem{ptau1}
M. Tanaka and R. Watanabe,
\prd {\bf82}, 034027 (2010) [arXiv:1005.4306 [hep-ph]].

\bibitem{Hsiao:2016pml1}
Y.~K.~Hsiao and C.~Q.~Geng, Chin. Phys. C \textbf{41}, 013101 (2017) [arXiv:1607.02718 [hep-ph]].

\bibitem{Kiselev:2002vz}
V.~V.~Kiselev [arXiv:hep-ph/0211021].

\bibitem{CHEN49}
C. H. Chang and Y. Q. Chen, Phys. Rev. D \textbf{49}, 3399 (1994).

\bibitem{Faustov:2022ybm}
R.~N.~Faustov, V.~O.~Galkin and X.~W.~Kang, Phys. Rev. D \textbf{106},  013004 (2022) [arXiv:2206.10277 [hep-ph]].

\bibitem{Wang:2008xt}
W.~Wang, Y.~L.~Shen and C.~D.~Lu, Phys. Rev. D \textbf{79}, 054012 (2009) [arXiv:0811.3748 [hep-ph]].

\bibitem{Nayak:2021djn}
L.~Nayak, S.~Patnaik, P.~C.~Dash, S.~Kar and N.~Barik, Phys. Rev. D \textbf{104}, 036012 (2021) [arXiv:2106.09463 [hep-ph]].

\bibitem{Huang:2018nnq}
Z.~R.~Huang, Y.~Li, C.~D.~Lu, M.~A.~Paracha and C.~Wang, Phys. Rev. D \textbf{98}, 095018 (2018) [arXiv:1808.03565 [hep-ph]].
\end{thebibliography}
\end{document}